\documentclass[11pt,fullpage]{article}

\usepackage{times,amsmath,epsfig}
\usepackage{graphicx}
\usepackage{placeins}
\usepackage{url}
\usepackage{subfig}
\usepackage{paralist}
\usepackage{caption}
\usepackage{multirow}
\usepackage{fullpage}

\begin{document}
\date{May 2013}

\title{Formal Representation of the SS-DB Benchmark and Experimental Evaluation in EXTASCID}

\author{
% 1st. author
Yu Cheng \hspace*{2cm} Florin Rusu\\
       \small{University of California, Merced}\\
       \small{5200 N Lake Road}\\
       \small{Merced, CA 95343}\\
       \small\texttt{\{ycheng4,frusu\}@ucmerced.edu}
}

\maketitle

%%%%%%%%%%%%%%%%%%%%%%%%%%%%%%%%%%%%%%%%%
%%%%%%%%%%%%%%%%%%%%%%%%%%%%%%%%%%%%%%%%%
%\input{mydefs}
\newcommand{\eat}[1]{}

\newcommand{\Ident}[1]{\mathcal{I}(#1)} % identity function

\newcommand{\ip}{{i^\prime}}
\newcommand{\Si}{\sum_{i\in I}}
\newcommand{\Sip}{\sum_{\ip \in I}}
\newcommand{\Sipni}{\sum_{\ip \in I, \ip\neq i}}
\newcommand{\Ip}{I^\prime}
\newcommand{\Il}{I_l}

\newcommand{\jp}{j^\prime}
\newcommand{\Sj}{\sum_{j\in I}}
\newcommand{\Sjp}{\sum_{\jp \in J}}
\newcommand{\Jp}{J^\prime}
\newcommand{\Jk}{J_k}

\newcommand{\F}{F}
\newcommand{\Fl}{F_l}
\newcommand{\ffi}{f_i} % conflict with \fi
\newcommand{\fip}{f_\ip}
\newcommand{\fav}{\overline{f}}
\newcommand{\flav}{\overline{f}_l}
\newcommand{\fSi}{\tilde{f}_i}
\newcommand{\fsi}{f_{\sigma(i)}}
\newcommand{\fgi}{f_{\gamma(i)}}
\newcommand{\fgip}{f_{\gamma(\ip)}}

\newcommand{\G}{G}
\newcommand{\Gl}{G_l}
\newcommand{\gi}{g_i}
\newcommand{\gip}{g_\ip}
\newcommand{\gj}{g_j}
\newcommand{\gk}{g_k}
\newcommand{\gav}{\overline{g}}
\newcommand{\glav}{\overline{g}_l}
\newcommand{\gkav}{\overline{g}_k}
\newcommand{\gSi}{\tilde{g}_i}
\newcommand{\gsi}{g_{\sigma(i)}}
\newcommand{\gsip}{g_{\sigma(\ip)}}

\newcommand{\gij}{g_{ij}}
\newcommand{\gipjp}{g_{\ip\jp}}
\newcommand{\glkav}{\overline{g}_{lk}}

\newcommand{\Hr}{H} % conflict with \H as Gamma
\newcommand{\Hk}{H_k}
\newcommand{\hj}{h_j}
\newcommand{\hjp}{h_\jp}
\newcommand{\hav}{\overline{h}}
\newcommand{\hkav}{\overline{h}_k}

\newcommand{\Sl}{\sum_{l=1}^n}
\newcommand{\Sk}{\sum_{k=1}^m}

\newcommand{\Pb}[1]{P\left[#1\right]}
\newcommand{\E}[1]{E\left[#1\right]}
\newcommand{\Var}[1]{\text{Var}\left(#1\right)}
\newcommand{\Cov}[2]{\text{Cov}(#1,#2)}
 
\newcommand{\Ps}[1]{P_\sigma\left[#1\right]}
\newcommand{\Es}[1]{E_\sigma\left[#1\right]}
\newcommand{\Eg}[1]{E_\gamma\left[#1\right]}
\newcommand{\Egs}[1]{E_{\gamma,\sigma}\left[#1\right]}
\newcommand{\SJ}[1]{\text{SJ}(#1)}
\newcommand{\Vars}[1]{\text{Var}_\sigma(#1)}
\newcommand{\Covs}[2]{\text{Cov}_\sigma(#1,#2)}
\newcommand{\Varg}[1]{\text{Var}_\gamma(#1)}
\newcommand{\Covg}[2]{\text{Cov}_\gamma(#1,#2)}
\newcommand{\Vargs}[1]{\text{Var}_{\gamma,\sigma}(#1)}
\newcommand{\Covgs}[2]{\text{Cov}_{\gamma,\sigma}(#1,#2)}
\newcommand{\N}[1]{\mathcal{N}(#1)}
\newcommand{\SErr}[1]{\text{SqErr}(#1)}
\newcommand{\RSErr}[1]{\sqrt{\SErr{#1}}}
\newcommand{\Err}[1]{\text{Err}(#1)}

\newcommand{\EstVar}[1]{\text{Est}_{#1}}

\newcommand{\f}{\mathcal{F}} % function symbol
\newcommand{\SUM}{\text{SUM}}
\newcommand{\COUNT}{\text{COUNT}}
\newcommand{\AVG}{\text{AVG}}
\newcommand{\STD}{\text{STD}}
\newcommand{\MIN}{\text{MIN}}
\newcommand{\MAX}{\text{MAX}}

\newcommand{\sm}[1]{\!#1\!} % reduces the space arround symbols

%alte definitii
\newtheorem{ex}{Example}
\newtheorem{thm}{Theorem}
\newtheorem{dfn}{Definition}
\newtheorem{prop}{Proposition}
\newtheorem{lema}{Lemma}
\newtheorem{corolar}{Corollary}
\newtheorem{aplicatie}{Application}

\newcommand{\Z}{\mathbb Z}

%\newcommand{\algorithmicinput}{\textbf{Input:}}
%\newcommand{\algorithmicoutput}{\textbf{Output:}}
%\newcommand{\algorithmicforeach}{\textbf{for each}}

% command for puting three figures on a line in ACM format
\newcommand{\threefiguresX}[9]{
\begin{figure*}[t]
\begin{center}
\begin{minipage}[htbp] {0.32\textwidth}
  \centering \includegraphics[width=\textwidth]{#1}
\caption{\small #2}\label{#3}
\end{minipage}
\hfil
\begin{minipage}[htbp] {0.32\textwidth}
  \centering \includegraphics[width=\textwidth]{#4}
\caption{\small #5}\label{#6}
\end{minipage}
\hfil
\begin{minipage}[htbp] {0.32\textwidth}
  \centering \includegraphics[width=\textwidth]{#7}
\caption{\small #8}\label{#9}
\end{minipage}
\end{center}
%\vspace*{-0.2in}
\end{figure*}
}

%%%%%%%%%%%%%%%%%%%%%%%%%%%%%%%%%%%%%%%%%
%%%%%%%%%%%%%%%%%%%%%%%%%%%%%%%%%%%%%%%%%
%\input{abstract}
\begin{abstract}

Evaluating the performance of scientific data processing systems is a difficult task considering the plethora of application-specific solutions available in this landscape and the lack of a generally-accepted benchmark. The dual structure of scientific data coupled with the complex nature of processing complicate the evaluation procedure further. SS-DB is the first attempt to define a general benchmark for complex scientific processing over raw and derived data. It fails to draw sufficient attention though because of the ambiguous plain language specification and the extraordinary SciDB results. In this paper, we remedy the shortcomings of the original SS-DB specification by providing a formal representation in terms of ArrayQL algebra operators and ArrayQL/SciQL constructs. These are the first formal representations of the SS-DB benchmark. Starting from the formal representation, we give a reference implementation and present benchmark results in EXTASCID, a novel system for scientific data processing. EXTASCID is complete in providing native support both for array and relational data and extensible in executing any user code inside the system by the means of a configurable metaoperator. These features result in an order of magnitude improvement over SciDB at data loading, extracting derived data, and operations over derived data.

\end{abstract}

%%%%%%%%%%%%%%%%%%%%%%%%%%%%%%%%%%%%%%%%%
%%%%%%%%%%%%%%%%%%%%%%%%%%%%%%%%%%%%%%%%%
%\input{introduction}
\section{Introduction}\label{sec:intro}

%%%%%%%%%%%%%%%%%%%%%%%%%%%
%problem
Scientific investigation represents an important source of Big Data. Science generate massive amounts of data through high-rate measurements of physical conditions, environmental and astronomical observations, and high-precision simulations of physical phenomena. While effectively storing the data is a challenge in itself, the main problem scientists face is how to efficiently process data in order to obtain novel insights and gain knowledge. Considering the plethora of application-specific solutions available in the scientific data processing landscape, selecting the optimal solution for a given problem is a challenging task. The lack of standardized benchmarks that allow for a principled evaluation of the available alternatives makes the selection process even more difficult.

%%%%%%%%%%%%%%%%%%%%%%%%%%%
%SS-DB
The \textit{Standard Science DBMS Benchmark (SS-DB)}~\cite{ssdb} is a recent attempt to create a general benchmark for the evaluation of scientific data processing systems. Similar to other popular benchmarks, e.g., the TPC benchmark suite~\cite{tpc}, SS-DB is modeled after a real application workload based on a complete workflow for processing astronomical images. Nonetheless, the benchmark operations are representative for a large class of scientific data manipulations. Unlike the Sloan Digital Sky Survey~\cite{sdss:article} which is targeted at a specific aspect in the scientific processing pipeline, the SS-DB benchmark encompasses a full spectrum of operations over raw and derived data. While this is an important step toward generality, it also introduces some serious problems. The uttermost limitation which hinders a broader benchmark implementation is that operations are expressed in plain language. There is no formal representation for the benchmark operations. The lack of a generally accepted formalism to represent multi-dimensional array operations -- the main component of the benchmark -- is a valid argument in this respect. The immediate effect is nevertheless negative -- we are aware of only two implementations of the benchmark, both presented in~\cite{ssdb} -- since lack of formalization makes impossible the definition of reference implementations. Other factors that drive the community away are the original results published in~\cite{ssdb} and the evolving state of SciDB~\cite{scidb}---the reference system for the benchmark. The original benchmark results compare SciDB to a relational-based implementation on top of MySQL database. The difference between the two systems is enormous -- 1 to 3 orders of magnitude -- in favor of SciDB, mostly due to architectural differences and the inefficient mapping of arrays on top of relations in MySQL. The extraordinary SciDB performance discourages others from implementing the benchmark. Moreover, SciDB is only a prototype suffering considerable modifications from one version to another. These propagate to frequent modifications to the SciDB implementation of the benchmark resulting in frequent updates to the reference benchmark results.

%%%%%%%%%%%%%%%%%%%%%%%%%%%
%EXTASCID
As illustrated by the SS-DB benchmark, scientific data have dual structure. Raw data are ordered multi-dimensional arrays while derived data are best represented as unordered relations. At the same time, scientific data processing requires complex operations over arrays and relations. These operations cannot be expressed using only standard linear and relational algebra operators, respectively. Existing scientific data processing systems address only a subset of these requirements. They are typically designed for a single data model, e.g., multi-dimensional arrays in SciDB, or they can handle complex processing only at the application level.

\textit{EXTASCID (EXTensible system for Analyzing SCIentific Data)} on the other hand is a complete and extensible system for scientific data processing. It supports natively both arrays as well as relational data. Complex processing is handled by a metaoperator that can execute any user code. EXTASCID provides unlimited extensibility by making the execution of arbitrary user code a central part of its design through the well-established User-Defined Aggregate (UDA) mechanism. As a result, EXTASCID supports in-database processing of full scientific workflows over both raw and derived data. Given all these desirable features provided in EXTASCID and the generality of the SS-DB benchmark, it is natural to ask what is the performance of EXTASCID on the SS-DB benchmark? Can all the complex benchmark operations be executed inside EXTASCID without moving data in the application layer? And how does the performance compare to SciDB?

%%%%%%%%%%%%%%%%%%%%%%%%%%%
%solutions & our contributions (high-level and detailed)
In this paper, we address the aforementioned shortcomings of the SS-DB benchmark and answer the questions on the generality and performance of EXTASCID. Our end goal is to propose a sound formal representation for the benchmark operations together with a reference EXTASCID implementation and reference results. On one hand, this strengthens dramatically the relevance of the benchmark and enforces its position as the reference benchmark for scientific data processing. Given that no alternatives exist, a broad acceptance of the SS-DB benchmark fills an important void in the evaluation of a large class of Big Data applications. On the other hand, the EXTASCID implementation of the benchmark provides another reference point in the evaluation of scientific data processing systems. The results prove that the integrated EXTASCID architecture supporting natively both arrays and relations is more suited for complex scientific processing over raw and derived data requiring a high degree of extensibility. To this end, our contributions can be summarized as follows:
\begin{compactitem}
\item We give a formal representation of the SS-DB benchmark in terms of ArrayQL algebra operators~\cite{aql:algebra}. We provide statements for the benchmark queries in the ArrayQL~\cite{aql:syntax} and SciQL~\cite{SciQL-ideas} query languages. These are the first formal representations of the SS-DB benchmark. They can be used as reference for implementation in other systems.
\item We present the design and implementation of EXTASCID---a novel system for scientific data processing. EXTASCID is complete in providing native support both for array and relational data and extensible in executing any user code inside the system by the means of a configurable metaoperator.
\item We provide a reference SS-DB implementation in EXTASCID starting from the formal representation of the benchmark in ArrayQL algebra.
\item We present results obtained by executing the SS-DB benchmark in EXTASCID. These are only the third reported results in the short history of the benchmark. When compared to the SciDB reference results, EXTASCID provides an order of magnitude improvement at data loading, extracting derived data, and operations over derived data. This is impressive considering the initial performance gap between SciDB and other scientific data processing systems.
\end{compactitem}

%%%%%%%%%%%%%%%%%%%%%%%%%%%
%roadmap
The rest of the paper is organized as follows. Section~\ref{sec:query-lang} presents array algebra formalisms and array query languages used to represent the SS-DB benchmark operations which are described in detail in Section~\ref{sec:ssdb}. The EXTASCID design and implementation are introduced in Section~\ref{sec:extascid}. The SS-DB implementation in EXTASCID starting from the formal array algebra representation is given in Section~\ref{sec:bench-implement} while the benchmark results and the comparison with SciDB are presented in Section~\ref{sec:results}. Related work is presented in Section~\ref{sec:rel-work}. We conclude in Section~\ref{sec:conclusions}.

%%%%%%%%%%%%%%%%%%%%%%%%%%%%%%%%%%%%%%%%%
%%%%%%%%%%%%%%%%%%%%%%%%%%%%%%%%%%%%%%%%%
%\input{query-lang}
\section{Array Query Languages}\label{sec:query-lang}

In order to analyze the SS-DB benchmark specification, we have to represent the benchmark operations in a formal query language. While relational algebra and SQL are standard formalisms for unordered relational data, there is no such algebra or query language commonly accepted for ordered array data. As a result, we settle for ArrayQL algebra~\cite{aql:algebra} and ArrayQL~\cite{aql:syntax} as our array algebra and query language, respectively. There are two reasons for our choice. First, these two formalisms are the most recent proposed in the literature. And second, they are part of the SciDB ecosystem~\cite{scidb:ssdbm-11}, similar to SS-DB. We discuss alternative array algebra formulations and query languages in the related work.

%%%%%%%%%%%%%%%%%%%%%%%%%%
\subsection{ArrayQL Algebra}\label{ssec:lang:AQLA}

Arrays are formalized as 3-tuples of the form \texttt{(box, valid, content)}, where \texttt{box} represents the domain of the array with fixed bounds on all dimensions, \texttt{valid} is a boolean map indicating which cells have valid values, and \texttt{content} is a function providing the values for the array cells. This is the first algebra that represents cell validity explicitly. The benefit is that both dense and sparse arrays can be formalized within the same algebra constructs.

Given the representation of an array as a 3-tuple, a new array is created by each operator, with a corresponding new 3-tuple. Operators define mappings between the original 3-tuple components and the new components. Without going into details, we present the most important operators defined in ArrayQL algebra in the following:
\begin{compactitem}
\item \texttt{SHIFT} array origin to a new position by changing the domain of the array components accordingly. It is useful when moving between coordinate systems.
\item \texttt{REBOX} changes the dimension sizes. It can either clip or extend the array domain. \texttt{REBOX} implements subsampling or range queries over dimensions, one of the most important array operations.
\item \texttt{FILTER} invalidates some array cells based on a content-only predicate. It is the direct equivalent of selection from relational algebra.
\item \texttt{FILL} transforms all the invalid cells to valid and assigns them a default value. Essentially, \texttt{FILL} transforms a sparse array into a dense one.
\item \texttt{APPLY} applies a function to each valid cell of an array.
\item \texttt{COMBINE} combines the content of two arrays having the same shape, but not necessarily the same validity. The content of the new array is computed by a function over the content of the argument arrays.
\item \texttt{INNERDJOIN} and \texttt{INNEREJOIN} are join operators over dimensions, and dimensions and attributes, respectively. Their semantics is equivalent to the natural join operator in relational algebra.
\item \texttt{REDUCE} generates a reduced version of an array by aggregating over one or more dimensions. Supported aggregate functions include the standard SQL aggregates.
\end{compactitem}

While ArrayQL algebra allows for a large variety of array operations to be expressed, there is an important feature that is completely missing from the algebra. This is the notion of adjacency or cell neighborhood. There is no operator that allows for aggregate functions to be applied to multiple adjacent cells centered on all the valid cells in the original array. While this operation can be expressed as a series of \texttt{REBOX} and \texttt{REDUCE} operators applied at all the cells in the original array, we argue that it is common enough to deserve an operator by itself.

An operator that handles adjacency is \texttt{APPLY} from AML~\cite{marathe:arrays}---this is a generalization of \texttt{APPLY} from ArrayQL algebra. We name this operator \texttt{APPLY+} to avoid confusion. The argument function is defined over an array shape and generates as output another shape. It is applied to every cell in the input array---shape centered on each cell, to be precise. The most common case is when the output array has exactly the same shape as the input array. In this case, the output shape is a single array cell and there is direct correspondence between the origin cell in the input array and the output cell. The main difference from \texttt{APPLY} is that the value of the output cell is a function of multiple adjacent cells in the input array. Moreover, \texttt{APPLY+} can specify which cells in the input array are considered as origin cells. In this situation, the output shapes are concatenated following the order in the input array to generate a dense array.

%%%%%%%%%%%%%%%%%%%%%%%%%%
\subsection{ArrayQL}\label{ssec:lang:AQL}

ArrayQL is an array creation and query language based on ArrayQL algebra. It is highly reminiscent of SQL and contains only two statements---CREATE ARRAY to create arrays at the schema level and SELECT FROM to query arrays. ArrayQL queries take as input arrays. The output can be either a new array -- with dimensions specified explicitly in the query as brackets -- or a relation---without any ordering constraint. Ranges on dimensions can be specified both for the input and the output arrays. In the case of input arrays, ranges correspond to sub-arrays, while in the case of the result array, ranges implement the \texttt{SHIFT} operator. If no ranges are provided, the complete dimension ranges of the input array(s) are automatically inherited. Structural joins between two arrays are specified by enumerating the arrays in the FROM clause and matching the dimension names. Overall, algebra operators are mostly implemented through index mappings. Not all ArrayQL algebra operators are specified in the language though. And not all the operations possible in the language by means of intricate index mappings are part of ArrayQL algebra.

SciQL~\cite{SciQL-ideas} is a direct precursor of ArrayQL with almost identical syntax. It has a very important feature not present in ArrayQL though. It supports the \texttt{APPLY+} operator as structural grouping. Thus, whenever the benchmark operations require it, we use SciQL queries instead of the less expressive ArrayQL.

%%%%%%%%%%%%%%%%%%%%%%%%%%%%%%%%%%%%%%%%%
%%%%%%%%%%%%%%%%%%%%%%%%%%%%%%%%%%%%%%%%%
%\input{ss-db}
\section{SS-DB Benchmark}\label{sec:ssdb}

The SS-DB benchmark~\cite{ssdb} is modeled based upon a real workflow for processing astronomical images. Although application-specific, SS-DB includes a full spectrum of operations over raw and derived data representative across various scientific domains. Queries in SS-DB are on 1-D arrays (e.g., polygon boundaries), dense and sparse 2-D arrays (e.g., images and astrophysical objects), and 3-D arrays (e.g., trajectories in space and time). Raw data ingestion and the computation of derived data are also part of the benchmark. In the following, we provide a detailed description of the SS-DB benchmark components and operations. Our contribution is to provide a formal representation for the benchmark operations based on the ArrayQL algebra and query language presented in Section~\ref{sec:query-lang}---the original specification in~\cite{ssdb} is in plain language. The abstract benchmark representation simplifies the understanding considerably and provides a clear specification for implementation in other systems.

%%%%%%%%%%%%%%%%%%%%%%%%%%%%%%%%%%%%%%%%%%%%%%%
\subsection{Raw Data}\label{ssec:ssdb:raw-data}

The basic data element is represented by a 2-D grid, i.e., dense array, corresponding to a sky image. The default size of the grid is a configurable parameter. The origin of the grid lies on a large 2-D plane $\left(10^{8} \text{ X } 10^{8}\right)$ corresponding to the entire sky. The origin has a higher chance to be placed towards the center of the domain than at other position in the space. This results in a dense region of grids lying in the central region of the domain and sparse everywhere else. Each grid cell contains $11$ integer values corresponding to a set of measurements taken at that point. The distribution of the values is chosen to reflect as close as possible real scientific data. An instance of the benchmark consists of multiple grids spread across the domain. They are grouped into cycles according to the time when the image was taken, thus introducing a third dimension. In essence, the complexity of the benchmark is given by the size of the grid and the number of cycles, with more and larger grids corresponding to more difficult benchmarks.

As a concrete example, consider the specifics of the normal SS-DB instance. The size of each grid is $7,500 \text{ X } 7,500$. There are a total of $400$ grids in the dataset, grouped into cycles of $20$, for a total of $20$ cycles. The overall size of the dataset is approximately $1\text{TB}$---each grid is $2.48\text{GB}$.

The ArrayQL definition for the raw dataset is:
\begin{equation}\label{eq:raw-def}
\begin{split}
& \texttt{CREATE ARRAY images (}\\
& \hspace*{0.5cm} \texttt{img\_id INTEGER DIMENSION [0:399],}\\
& \hspace*{0.5cm} \texttt{x INTEGER DIMENSION [0:7499],}\\
& \hspace*{0.5cm} \texttt{y INTEGER DIMENSION [0:7499],}\\
& \hspace*{0.5cm} \texttt{v1 INTEGER, $\dots$, v11 INTEGER}\\
& \texttt{)}
\end{split}
\end{equation}
It is important to notice that this representation is in the local coordinate system corresponding to each image. The origin of the images does not need to coincide. The origin of the grids in the global coordinate system is stored in the 1-D array \texttt{image\_origin}:
\begin{equation}\label{eq:orig-def}
\begin{split}
& \texttt{CREATE ARRAY image\_origin (}\\
& \hspace*{0.5cm} \texttt{img\_id INTEGER DIMENSION [0:399],}\\
& \hspace*{0.5cm} \texttt{x INTEGER, y INTEGER}\\
& \texttt{)}
\end{split}
\end{equation}
In the global coordinate system, \texttt{images} is a sparse array over the 3-D space $\left(400 \text{ X } 10^{8} \text{ X } 10^{8}\right)$ with a single $\left(7,500 \text{ X } 7,500\right)$ dense sub-array at each \texttt{img\_id} index. The distinction between these two representations is significant for query execution. Depending on which coordinate system is used, the \texttt{SHIFT} operator in ArrayQL algebra has to be applied to change the origin of the array prior to query processing.

%%%%%%%%%%%%%%%%%%%%%%%%%%%%%%%%%%%%%%%%%%%%%%%
\subsection{Derived Data}\label{ssec:ssdb:derived-data}

Raw data consist of dense grids with values associated to each cell in the domain. If we consider the entire 3-D space (\texttt{img}\_\texttt{id}-\texttt{x}-\texttt{y}) though, raw data are very sparse, i.e., only a fraction of $0.5625\times 10^{-8}$ cells have values. Derived data are generated from the raw data through clustering. There are two types of clustering specified in the benchmark. Cooking, or observation creation, is local clustering inside each grid based on cell values and cell neighborhood relationships. Grouping is distance-based clustering applied to the previously obtained observations across the grids in the same cycle. It is important to notice that derived data represent sparse arrays both in the 2-D domain (\texttt{x}-\texttt{y}) as well as in the 3-D space (\texttt{img}\_\texttt{id}-\texttt{x}-\texttt{y}).

%%%%%%%%%%%%%%%%%%%%%%%%%%%%%%%%%%%%%%%%%%%%%%%
\subsubsection{Observations}\label{ssec:ssdb:cooking}

Observations are extracted, i.e., "cooked", from raw data based on a user-defined function (UDF) over cell values. Intuitively, all the cells that are part of an observation satisfy two conditions: they are neighbors and the UDF holds for each individual cell. They form a cluster with a common property. As an example, consider adjacent pixels in an image with the R component in RGB having values greater than $100$. The actual number of observations in a grid is strictly determined by the parameters of the UDF. It is likely though that only a small number of cells are part of observations. To enforce this explicitly, the benchmark imposes two conditions. It limits the size of the bounding box and the number of edges in the boundary polygon corresponding to each observation. In addition to the data corresponding to each individual cell, a series of aggregated attributes are defined for an observation: the center, the bounding box and the boundary polygon, and some additional domain-specific properties.

To understand the semantics of the cooking operation, we examine how it can be expressed as an array algebra formula---a sequence of array algebra operators, to be precise. Abstractly, cooking corresponds to the labeling operation from image processing, i.e., identify all the groups of adjacent cells, i.e., observations, satisfying some common property. In this case, the property is that the value of one attribute, e.g., \texttt{v1}, is greater than a given threshold. The adjacent cells are defined as a hypercube of a configurable size and centered on each cell in the input array---the hypercube is statically specified at query time and is fixed for all the cells in the array. The result of cooking is an array with exactly the same shape and size. Cell values correspond to the unique identifier assigned to each observation. Other properties corresponding to the whole observation, e.g., center, bounding box, and boundary polygon, can also be computed once the observation is determined.

The sequence of array algebra operators -- ArrayQL algebra enhanced with the generalized \texttt{APPLY+} from AML -- that implement cooking is the following:
\begin{compactenum}
\item \texttt{FILTER} the array with the cell predicate. \texttt{valid} is set to true only for the cells satisfying the predicate.
\item Assign a unique id to each cell that is still valid. This can be done using a function of the array indexes. The id is an additional attribute to the original array.
\item \texttt{APPLY+} a function that sets each cell to the minimum id of all the neighbor cells. This is done for each cell in the input array. The result is a new array with the same id in the cells corresponding to an observation. To generate the entire observation, \texttt{APPLY+} has to be invoked iteratively until the source and result arrays are identical---steady state.
\end{compactenum}
Computing observation properties is not a straightforward algebra operation either because observations have arbitrary shapes. The following steps are executed for each observation, extracted iteratively by \texttt{FILTER} on the observation id:
\begin{compactenum}
\item \texttt{INNERDJOIN} is used to get the raw data corresponding to the cells in the observation.
\item \texttt{REDUCE} is applied to generate aggregate properties. These can be stored at all the cells that are part of the observation or only at a designated cell, e.g., the center.
\end{compactenum}

%%%%%%%%%%%%%%%%%%%%%%%%%%%%%%%%%%%%%%%%%%%%%%%%%%%%%%%%%%%%%%%%%%%%%%
\eat{
An equivalent expression to \texttt{APPLY} can be written in Array algebra~\cite{baumann:ngits} as the composition of two operators applied to every image in the dataset:
\begin{equation}\label{eq:AA:cook}
\begin{split}
\texttt{obs}_{\texttt{i+1}} =& \texttt{ MARRAY}_{\texttt{sdom}(\texttt{obs}_{\texttt{i}}), \texttt{x}} (\\
& \hspace{0.2cm}\texttt{ COND}_{\texttt{min, sdom(nhood), y}} \left( \texttt{obs}_{\texttt{i}} (\texttt{x+y}) \right))
\end{split}
\end{equation}
where $\texttt{obs}_{\texttt{0}}$ is initialized to the \texttt{images} array and \texttt{nhood} is the neighborhood shape to be considered for each array cell. In the case of cooking, \texttt{nhood} is defined as:

\begin{table}[h]\label{ex:cook:nhood}
    \begin{center}
      \begin{tabular}{|c|c|c|}

    \hline
		1 & 1 & 1\\
	\hline
		1 & 1 & 1\\
	\hline
		1 & 1 & 1\\
	\hline

      \end{tabular}
    \end{center}
\end{table}
}
%%%%%%%%%%%%%%%%%%%%%%%%%%%%%%%%%%%%%%%%%%%%%%%%%%%%%%%%%%%%%%%%%%%%%%

The SciQL query corresponding to \texttt{APPLY+} for the first image in the array \texttt{images} is:
\begin{equation}\label{eq:SciQL:cook}
\begin{split}
& \texttt{SELECT [x], [y], MIN(id)}\\
& \texttt{FROM images[0]}\\
& \texttt{GROUP BY images[0][x-1:x+1][y-1:y+1]}
\end{split}
\end{equation}
For cooking to work correctly, it is required that both the array algebra operator \texttt{APPLY+} and the structural grouping in SciQL can identify the valid array cells.

%%%%%%%%%%%%%%%%%%%%%%%%%%%%%%%%%%%%%%%%%%%%%%%
\subsubsection{Groups}\label{ssec:ssdb:grouping}

The second form of derived data consist of groups of observations. A group contains observations from different grids in the same cycle having the centers close to each other---the centers are not required to coincide. The actual definition of closeness is specified through a UDF. Intuitively, a group can be imagined as a cluster in the 3-D space of grid cycles. There is no requirement though that the observations need to be in adjacent grids -- as is the case for observations -- since the distance function already takes into account the distance across the time dimension, i.e., \texttt{img}\_\texttt{id}. The center and bounding box of a group are defined as in the case of observations, from the centers and the bounding boxes corresponding to member observations.

In order to write an array algebra expression for grouping, we need to get a better understanding of the operation. The important detail to remark is that the neighbors are determined according to a distance function rather than using a fixed shape centered on the observation center. Nonetheless, the neighbors can be represented as an irregular 3-D array computed based on a discretized version of the distance function. Thus, we can view grouping as a 3-D version of cooking with an irregular neighborhood shape operating on observation centers. Given an origin or reference observation, the neighborhood hypercube expands with the distance along the time dimension. The difficult part is to compute the discrete hypercube from the continuous distance function.

To compute the group corresponding to a reference observation, the following ArrayQL algebra operators have to be invoked:
\begin{compactenum}
\item \texttt{APPLY+} labeling -- set the observation id -- to all the observations in the reference observation neighborhood. In this case, labeling is done in a single shot, not iteratively.
\item \texttt{APPLY+} the same labeling as above for all the observations labeled before---they are part of the same group. This is done step-by-step along the time dimension.
\item Once the observations in a group are determined, \texttt{REDUCE} is called on the \texttt{INNERDJOIN} result with the observation data corresponding to all the observations in the group to compute aggregate properties for the group.
\end{compactenum}
The main difference between cooking and grouping is the irregular shape passed as argument to \texttt{APPLY+}. This difference is very significant in the case of SciQL though. Based on the examples given in~\cite{SciQL-ideas}, we argue that it is not possible to write grouping as a SciQL query because SciQL can handle only regular hypercubes.

%%%%%%%%%%%%%%%%%%%%%%%%%%%%%%%%%%%%%%%%%%%%%%%
\subsection{Queries}\label{ssec:ssdb:queries}

The benchmark defines a series of nine queries---three on raw data, three on observations, and three on groups. A general characteristic across all the queries is that instead of applying them to an entire grid, they typically operate on a slab of the space which is specified as part of the query. The size and position of the slab are important parameters that control the difficulty level of the benchmark. This operation is known as \textit{subsampling} or \textit{range} query and is highly dependent on the storage strategy. The ideal situation is to execute all the queries by reading only the required data and nothing extra. This is hard to enforce across all the possible ranges.

Before we proceed to provide the array algebra expression for each query in the benchmark, we discuss how to handle subsampling since it is embedded in almost all the benchmark queries. Subsampling can be expressed in ArrayQL algebra by the \texttt{REBOX} operator. \texttt{REBOX} clips the original array to the query slab by effectively reducing the size of its \texttt{box}. The position of the array origin changes accordingly. If the dimension sizes have to be preserved, a second call to \texttt{REBOX} can extend the array to its original size while invalidating the cells that are not part of the subsample. Intuitively, the same effect can be achieved by a single call to \texttt{FILTER} with the range conditions on dimensions. This is not supported in the current ArrayQL algebra~\cite{aql:algebra} since \texttt{FILTER} accepts conditions exclusively on the cell content and not on dimensions. We assume that in all the queries that require subsampling, \texttt{REBOX} is first applied to clip the array to the query range. Thus, our array algebra expressions do not represent \texttt{REBOX} explicitly unless necessary.

%%%%%%%%%%%%%%%%%%%%%%%%%%%%%%%%%%%%%%%%%%%%%%%
\subsubsection{Raw Data}\label{ssec:ssdb:query:raw}

%%%%%%%%%%%%%%%%%%%%%
\textbf{Q1 Aggregation.} "For the $20$ images in each cycle and for a slab of size $\left[\texttt{T1}, \texttt{U1}\right]$ in the local coordinate space, starting at $\left[\texttt{X1}, \texttt{Y1}\right]$, compute the average value of \texttt{vi} for a random value of \texttt{i}."~\cite{ssdb}

The ArrayQL algebra operators for the first cycle are:
\begin{equation}\label{eq:AQLA:Q1}
\begin{split}
& \texttt{R1 = REBOX(images, [img\_id=0:19,}\\
& \hspace*{1.2cm}\texttt{x=X1:X1+T1, y=Y1:Y1+U1])}\\
& \texttt{R2 = REDUCE(R1, \{img\_id, x, y\}, avg(vi))}
\end{split}
\end{equation}
Similar expressions can be written for the other cycles by changing the range on \texttt{img\_id} in the \texttt{REBOX} operator. While Q1 supports different aggregate operators, it can be generalized further by allowing induced functions~\cite{baumann:ngits} over attribute values inside the aggregate. This can be done by adding a call to the \texttt{APPLY} operator before \texttt{REDUCE}:
\begin{equation}\label{eq:AQLA:Q1-ind}
\texttt{APPLY(R1, f(v1, $\dots$, v11))}
\end{equation}

The ArrayQL syntax for Q1 is a simple SQL aggregation with the ranges specified after the array rather than in the \texttt{WHERE} clause:
\begin{equation}\label{eq:AQL:Q1}
\begin{split}
& \texttt{SELECT AVG(Vi)}\\
& \texttt{FROM images[0:19, X1:X1+T1, Y1:Y1+U1]}
\end{split}
\end{equation}

%%%%%%%%%%%%%%%%%%%%%
\textbf{Q2 Recooking.} "For a slab of size $\left[\texttt{T1}, \texttt{U1}\right]$ starting at $\left[\texttt{X1}, \texttt{Y1}\right]$, recook the raw imagery for the first image in the cycle with a different clustering function."~\cite{ssdb}

The same cooking process for computing observations is applied to a slab of an image and with a different filtering condition. These are marginal modifications to the sequence of array algebra operators presented in Section~\ref{ssec:ssdb:cooking}.

%%%%%%%%%%%%%%%%%%%%%
\textbf{Q3 Regridding.} "For a slab of size $\left[\texttt{T1}, \texttt{U1}\right]$ starting at $\left[\texttt{X1}, \texttt{Y1}\right]$, regrid the raw data for the images in the cycle, such that the cells collapse $10:3$. All the \texttt{vi} values in the raw data are regridded in this process by an interpolation function."~\cite{ssdb}

This operation modifies the size of the grid dimensions. While Q3 corresponds to image shrinking, it is equally possible to imagine a version that enlarges the grid with a specified ratio. The important aspect is that the number of cells reduces and the values in each new cell have to be determined accordingly. The standard solution is to compute the new value based on the values in adjacent cells using an interpolation function. Think of a mapping from a set of cells in the original grid to a single cell in the new grid, each making its share of contribution to the new value. In terms of array algebra operators, regridding is very similar to cooking, i.e., a neighborhood shape is applied at cell positions in the input array to generate cells in the output array. The main difference is that not all the cells in the input array are considered as origin in the case of regridding. How many and which is determined by the regrid ratio.

Out of all the array algebra formulations in the literature, only AML~\cite{marathe:arrays} supports cell selection through bit patterns. Thus, the same generalized \texttt{APPLY+} operator used for cooking can be also used for regridding. In addition to the neighborhood shape and the aggregate function, a bit pattern identifying the cells where \texttt{APPLY+} is invoked has to be specified. The bit pattern takes the form of a regular expression of 0's and 1's that is applied repetitively along the dimension domain. There is one such bit pattern for each dimension. \texttt{APPLY+} is invoked only at those cells where the bit pattern is 1 for all the dimensions. For Q3, the bit pattern on both \texttt{x} and \texttt{y} is the same:
\begin{equation}\label{eq:AQL:Q3-bit}
\texttt{(1001001000)}^{*}
\end{equation}
Since none of the array query languages in the literature implements the generalized \texttt{APPLY+} operator -- with the bit pattern selection -- we argue that Q3 cannot be expressed directly as a query---only as a UDF over the entire array.

%%%%%%%%%%%%%%%%%%%%%%%%%%%%%%%%%%%%%%%%%%%%%%%
\subsubsection{Observations}\label{ssec:ssdb:query:obs}

%%%%%%%%%%%%%%%%%%%%%
\textbf{Q4 Observation Aggregation.} "For the observations in the cycle with centers in a slab of size $\left[\texttt{T2}, \texttt{U2}\right]$ starting at $\left[\texttt{X2}, \texttt{Y2}\right]$ in the world coordinate space, compute the average value of observation attribute \texttt{oi}, for a randomly chosen \texttt{i}."~\cite{ssdb}

Q4 has exactly the same array algebra representation as Q1 if we consider observation centers to be the only valid cells of a sparse array \texttt{obs\_center}. The evaluation of the two queries is completely different though because \texttt{obs\_center} is a sparse array and the range condition is given in the global coordinate space. As a result, not every image has observations in the given range---there are images that do not even overlap the range. Q4 for the first cycle can be written in ArrayQL as follows:
\begin{equation}\label{eq:AQL:Q4}
\begin{split}
& \texttt{SELECT AVG(oi)}\\
& \texttt{FROM obs\_center[0:19,X2:X2+T2,Y2:Y2+U2]}
\end{split}
\end{equation}

%%%%%%%%%%%%%%%%%%%%%
\textbf{Q5 Polygons.} "For the observations in the cycle and for a slab of size $\left[\texttt{T2}, \texttt{U2}\right]$ starting at $\left[\texttt{X2}, \texttt{Y2}\right]$ in the world coordinate space, compute the observations whose polygons overlap the slab."~\cite{ssdb}

This query is similar to Q4 with the difference that instead of requiring the observation center to be contained in the slab, the query considers the observations for which the boundary polygon intersects with the slab. An alternative is to consider the bounding box instead of the polygon.

Consider observations to be represented as the only valid cells of a sparse array \texttt{obs}. The observation id is the single value stored in each cell that is part of an observation. The ArrayQL algebra expression for Q5 can then be written as:
\begin{equation}\label{eq:AQLA:Q5}
\begin{split}
& \texttt{R1 = REBOX(obs, [img\_id=0:19,}\\
& \hspace*{1.2cm}\texttt{x=X2:X2+T2, y=Y2:Y2+U2])}\\
& \texttt{R2 = REDUCE(R1, \{img\_id, x, y\},}\\
& \hspace*{1.2cm}\texttt{count distinct(obs\_id))}
\end{split}
\end{equation}
The ArrayQL syntax is identical to Eq.~(\ref{eq:AQL:Q1}). The only difference is that \texttt{COUNT DISTINCT} is used instead of \texttt{AVG}.

%%%%%%%%%%%%%%%%%%%%%
\textbf{Q6 Density.} "For the observations in the cycle and for a slab of size $\left[\texttt{T2}, \texttt{U2}\right]$ starting at $\left[\texttt{X2}, \texttt{Y2}\right]$ in the world coordinate space, group the observations spatially into \texttt{D4} by \texttt{D4} tiles, where each tile may be located at any integral coordinates within the slab. Find the tiles containing more than \texttt{D5} observations."~\cite{ssdb}

It is not clear from the query definition when an observation is considered to be part of a tile---if the observation center is contained in the tile or if any observation cell is part of the tile. We consider the first version and use the array corresponding to observation centers in the array algebra expressions. As with all the other queries that require access to neighboring cells, Q6 cannot be expressed using only ArrayQL algebra operators. The generalized \texttt{APPLY+} operator has to be used for grouping at every cell with a neighborhood shape $\texttt{D4} \times \texttt{D4}$ having the upper left corner at the cell. With this extension, the ArrayQL algebra representation is:
\begin{equation}\label{eq:AQLA:Q6}
\begin{split}
& \texttt{R1 = REBOX(obs\_center, [img\_id=0:19,}\\
& \hspace*{1.2cm}\texttt{cx=X2:X2+T2, cy=Y2:Y2+U2])}\\
& \texttt{R2 = APPLY+(R1[i], [1,D4,D4],}\\
& \hspace*{1.2cm}\texttt{count(center) AS density)}\\
& \texttt{R3 = FILTER(R2, density $\geq$ D5)}
\end{split}
\end{equation}
Notice that \texttt{R2} and \texttt{R3} are computed separately for each image $i=[0:19]$ in the cycle. In SciQL, they correspond to structural grouping followed by \texttt{HAVING}:
\begin{equation}\label{eq:SciQL:Q6}
\begin{split}
& \texttt{SELECT cx, cy, COUNT(center) AS density}\\
& \texttt{FROM obs\_center[i][X2:X2+T2][Y2:Y2+U2]}\\
& \texttt{GROUP BY obs\_center[1][cx:cx+D4][cy:cy+D4]}\\
& \texttt{HAVING density $\geq$ D5}
\end{split}
\end{equation}

%%%%%%%%%%%%%%%%%%%%%%%%%%%%%%%%%%%%%%%%%%%%%%%
\subsubsection{Groups}\label{ssec:ssdb:query:groups}

%%%%%%%%%%%%%%%%%%%%%
\textbf{Q7 Centroid.} "Find each group whose center falls in the slab of size $\left[\texttt{T2}, \texttt{U2}\right]$ starting at $\left[\texttt{X2}, \texttt{Y2}\right]$ in the world coordinate space at any time \texttt{t}. The center is defined to be the average value of the centers recorded for all the observations in the group."~\cite{ssdb}

This is a 3-D query over an array \texttt{group\_center} containing valid cells only for the group centers. One such array corresponds to each cycle. This array can be computed after the groups are determined. With this representation, the query for the first cycle can be expressed as a simple \texttt{REBOX} in ArrayQL algebra:
\begin{equation}\label{eq:AQLA:Q7}
\begin{split}
& \texttt{REBOX(group\_center, [cycle=0,}\\
& \hspace*{1.2cm} \texttt{cx=X2:X2+T2, cy=Y2:Y2+U2])}
\end{split}
\end{equation}
and in ArrayQL as:
\begin{equation}\label{eq:AQL:Q7}
\begin{split}
& \texttt{SELECT group\_id}\\
& \texttt{FROM group\_center[0,X2:X2+T2,Y2:Y2+U2]}
\end{split}
\end{equation}

%%%%%%%%%%%%%%%%%%%%%
\textbf{Q8 Center trajectory.} "Define trajectory to be the sequence of centers of the observations in an observation group. For each trajectory that intersects a slab of size $\left[\texttt{T3}, \texttt{U3}\right]$ starting at $\left[\texttt{X2}, \texttt{Y2}\right]$ in the world coordinate space, produce the raw data for a \texttt{D6} by \texttt{D6} tile centered on each center for all images that intersect the tile."~\cite{ssdb}

Q8 consists of two phases. First, the groups whose trajectory intersects a given slab are determined. Second, for each grid containing an observation in the group, the cells in a given tile centered on the group center in that grid are returned. It is important to notice that Q8 -- as well as Q9 -- requires access both to the raw images as well as to groups.

While in Q7 the center of a group is defined as a single 2-D point over all the observations in the group, in Q8 there is a center at each image that has observations in the group---there are at most $20$ centers for a group. They define the trajectory of a group as a 3-D array in the global coordinate space. Then the trajectories that intersect the query slab, i.e., at least one center is contained in the slab, can be extracted with a simple \texttt{REBOX} operator applied to the corresponding \texttt{group\_center\_img} array:
\begin{equation}\label{eq:AQLA:Q8-1}
\begin{split}
& \texttt{R1 = REBOX(group\_center\_img,}\\
& \hspace*{1.2cm}\texttt{[cycle=0, img\_id=0:19,}\\
& \hspace*{1.2cm}\texttt{ cx=X2:X2+T3, cy=Y2:Y2+U3])}
\end{split}
\end{equation}
For each valid cell in \texttt{R1}, raw data have to be extracted from the corresponding image. This can be done with \texttt{REBOX} on the image representation in the global coordinate space. Assuming we operate on the first image in the cycle at observation center $\texttt{<ocx, ocy>}$, the ArrayQL algebra operator sequence is:
\begin{equation}\label{eq:AQLA:Q8-2}
\begin{split}
& \texttt{R2 = SHIFT(images[0], <orig\_x, orig\_y>)}\\
& \texttt{R3 = REBOX(images[0],}\\
& \hspace*{1.2cm}\texttt{[ocx-D6:ocx+D6, ocy-D6:ocy+D6])}
\end{split}
\end{equation}
\texttt{<orig\_x, orig\_y>} is the image origin in the global coordinate system as stored in \texttt{image\_origin} (Eq.~\ref{eq:orig-def}).
Since the conversion of \texttt{REBOX} to ArrayQL is standard (see Eq. (\ref{eq:AQLA:Q7}) and~(\ref{eq:AQL:Q7})), we do not include it here.

%%%%%%%%%%%%%%%%%%%%%
\textbf{Q9 Polygon trajectory.} "Define trajectory to be the sequence of polygons that correspond to the boundary of the observation group. For a slab $\left[\texttt{T3}, \texttt{U3}\right]$ starting at $\left[\texttt{X2}, \texttt{Y2}\right]$, find the groups whose trajectory overlaps the slab at some time \texttt{t} and produce the raw data for a \texttt{D6} by \texttt{D6} tile centered on each center for all images that intersect the slab."~\cite{ssdb}

Q9 is similar to Q8. The only difference is the array used in \texttt{REBOX} over the derived data. The original observations array \texttt{obs} containing the groups the observation is part of and shifted to the origin in the global coordinate space has to be used instead of the \texttt{group\_center\_img} array. This array contains valid cells for all the points that are part of observations. It allows us to identify the groups overlapping the query slab and their centers at each position in the cycle.

%%%%%%%%%%%%%%%%%%%%%%%%%%%%%%%%%%%%%%%%%%%%%%%
\subsection{Discussion}\label{ssec:ssdb:discuss}

Given the algebra representation for all the operations in the benchmark, we analyze what array structures and operators are required for an actual implementation. The 3-D array corresponding to the raw images is given in Eq. (\ref{eq:raw-def}). \texttt{images} is represented in the local coordinate system. This representation suffices to answer Q1--Q3 and to execute the cooking. For Q8 and Q9, the representation of \texttt{images} in the global coordinate system is required. This can be obtained by shifting the origin of the images based on \texttt{image\_origin}. For grouping, only the observation centers are required. A sparse 3-D array \texttt{obs\_center} over the global coordinates containing valid cells only where observation centers are located can do the job. The data inside the cell contain the observation id and a set of attributes (Q4, Q6). All the cells that are part of observations are required for Q5. They can be represented as a sparse 3-D array \texttt{obs} with the observation id as the single attribute. This array is also required to answer Q9. Group centers are required to answer Q7 and Q8. In Q7, the single group centers can be represented as a sparse 3-D array \texttt{group\_center} containing the group ids in the valid cells. In Q8, group centers are recorded for each image in the cycle---they can be different from one image to another. Thus, a sparse 4-D array \texttt{group\_center\_img} with the group ids stored in the valid cells is required. Notice that cycle represents the additional dimension in the arrays corresponding to groups. Table~\ref{tbl:bench-arrays} summarizes the arrays required to answer the benchmark queries.

%%%%%%%%%%%%%%%%%%%%%
\begin{table}[htbp]
    \begin{center}
      \begin{tabular}{|l|l|l|}

    \hline
	\textbf{Name} & \textbf{Type} & \textbf{Coordinates} \\

	\hline\hline
	\texttt{images} & 3-D grid & local \\
	\hline
	\texttt{image\_origin} & 1-D grid & global \\
	\hline
	\texttt{obs} & 3-D sparse & local \\
	\hline
	\texttt{obs\_center} & 3-D sparse & global \\
	\hline
	\texttt{group\_center} & 3-D sparse & global \\
	\hline
	\texttt{group\_center\_img} & 4-D sparse & global \\
	\hline

      \end{tabular}
    \end{center}

    \caption{Arrays used in the SS-DB benchmark.}
    \label{tbl:bench-arrays}
\end{table}
%%%%%%%%%%%%%%%%%%%%%

In terms of the ArrayQL algebra operators, \texttt{REBOX} is used in all the range queries. The generalized form of \texttt{APPLY+} that is present only in AML is used for cooking, grouping, Q2, Q3, and Q6. The calls differ in terms of the cells where \texttt{APPLY+} is invoked, the shape of the neighborhood, and the aggregator. \texttt{REDUCE} is applied whenever an aggregate has to be computed, e.g., Q1 and Q4. \texttt{FILTER}, \texttt{SHIFT}, and \texttt{INNERDJOIN} are the other algebra operators used throughout the benchmark operations. They are all summarized in Table~\ref{tbl:bench-operators}.

%%%%%%%%%%%%%%%%%%%%%
\begin{table}[htbp]
    \begin{center}
      \begin{tabular}{|l|l|}

    \hline
	\textbf{Query} & \textbf{Algebra operators} \\

	\hline\hline
	Cooking, Q2 & \texttt{FILTER}, \texttt{APPLY+}, \texttt{INNERDJOIN}, \texttt{REDUCE} \\
	\hline
	Grouping & \texttt{APPLY+}, \texttt{INNERDJOIN}, \texttt{REDUCE} \\
	\hline
	Q1, Q4, Q5 & \texttt{REBOX}, \texttt{REDUCE} \\
	\hline
	Q3 & \texttt{APPLY+} \\
	\hline
	Q6 & \texttt{REBOX}, \texttt{APPLY+}, \texttt{FILTER} \\
	\hline
	Q7 & \texttt{REBOX} \\
	\hline
	Q8, Q9 & \texttt{REBOX}, \texttt{SHIFT} \\
	\hline

      \end{tabular}
    \end{center}

    \caption{Algebra operators used in the SS-DB benchmark.}
    \label{tbl:bench-operators}
\end{table}
%%%%%%%%%%%%%%%%%%%%%

%%%%%%%%%%%%%%%%%%%%%%%%%%%%%%%%%%%%%%%%%
%%%%%%%%%%%%%%%%%%%%%%%%%%%%%%%%%%%%%%%%%
%\input{extascid}
\section{EXTASCID}\label{sec:extascid}

EXTASCID is built around the massively parallel GLADE~\cite{glade:sigmod} architecture for data aggregation. While it inherits the extensibility provided by the original Generalized Linear Aggregate (GLA)~\cite{glade:sigmod} interface implemented in GLADE, EXTASCID enhances this interface considerably with functions specific to scientific processing. This requires significant extensions to the GLADE execution strategy in order to provide additional flexibility and to optimize array processing. The design of the EXTASCID parallel storage manager with native support for relations and arrays is entirely novel---GLADE works only for relational data.

Given its descent from GLADE, EXTASCID also satisfies the standard requirements for scientific data processing---support for massive datasets and parallel processing. Contrary to existent scientific data processing systems designed for a target architecture, typically shared-nothing, EXTASCID is \textit{architecture-independent}. It runs optimally both on shared-memory, shared-disk servers as well as on shared-nothing clusters. The reason for this is the exclusive use of thread-level parallelism inside a processing node while process-level parallelism is used only across nodes.

%%%%%%%%%%%%%%%%%%%%%
\begin{figure}[htbp]
\centering
\includegraphics[width=0.8\textwidth]{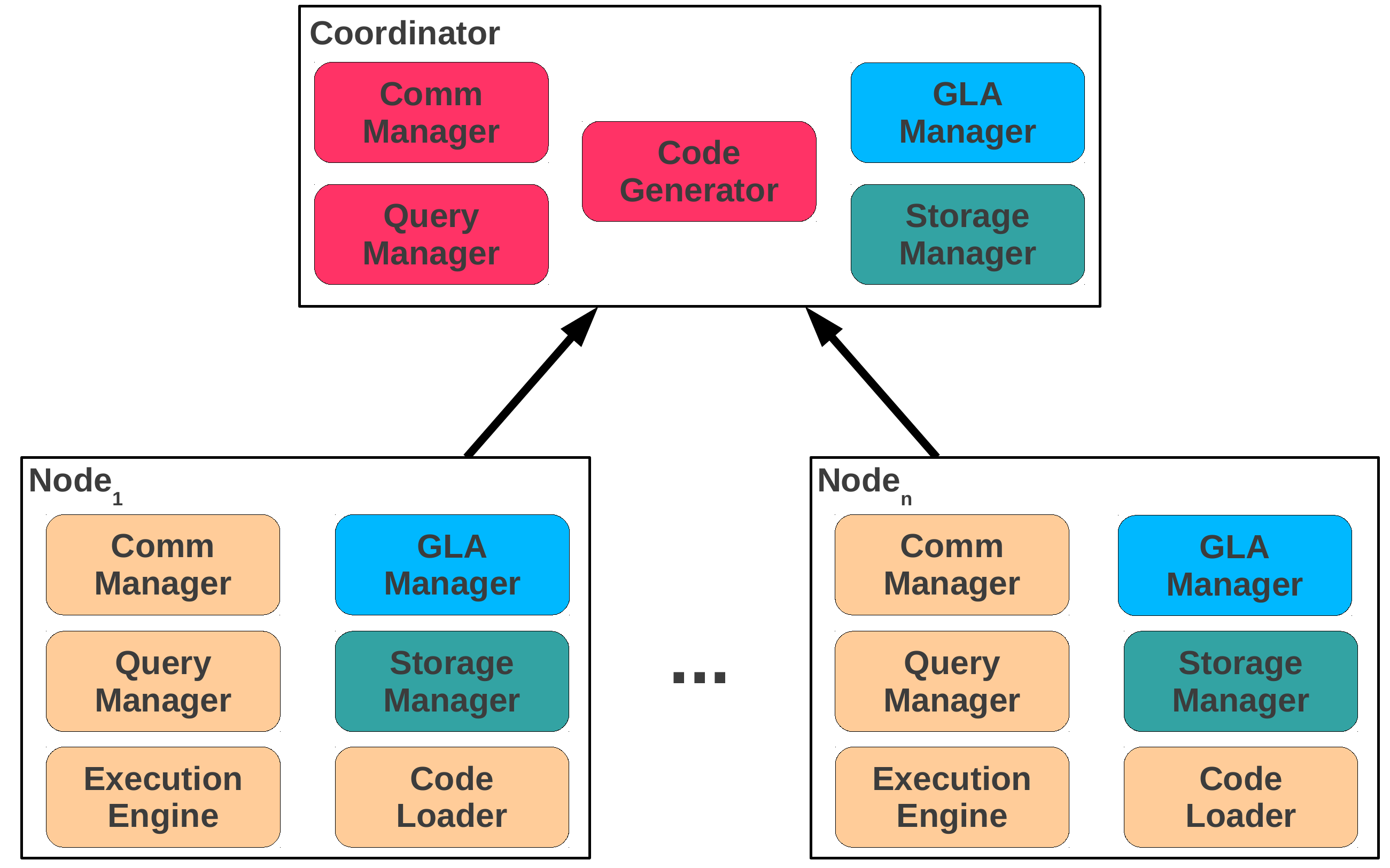}
\caption{EXTASCID system architecture.}
\label{fig:architecture}
\end{figure}
%%%%%%%%%%%%%%%%%%%%%

In a nutshell, EXTASCID is a parallel data processing system that executes any computation specified as a GLA using a merge-oriented execution strategy supported by a push-based storage manager. The storage manager is designed with special consideration for multi-dimensional range-based data partitioning in order to support efficient array processing. To allow for wide extensibility in terms of the supported user code and to extract maximum performance, GLAs are dynamically compiled inside EXTASCID at runtime following the optimized code generation mechanism proposed in the DataPath system~\cite{datapath}.

As shown in Figure~\ref{fig:architecture}, EXTASCID consists of two types of entities: a coordinator and one or more executor processes. The coordinator is the interface between the user and the system. Since it does not manage any data except the catalog metadata, the coordinator does not execute any data processing task. These are the responsibility of the executors, typically one for each physical processing node. It is important to notice that the executors act as completely independent entities, in charge of their data and of the physical resources. The coordinator as well as the executors consist of multiple components, depicted in Figure~\ref{fig:architecture}. While the components are inherited from the GLADE~\cite{glade:sigmod} architecture, significant changes are required in order to support array storage and processing in addition to the native relational data model. We discuss the changes to the two most important components -- Storage Manager and GLA Manager -- in separate sections. In the following, we summarize the main functionality of the other components.

\textit{Query Manager}. The query manager at the coordinator is responsible for setting-up and managing the processing requested by the user. In the case of multiple executor processes, the query manager builds the aggregation tree used for merging the GLAs.

\textit{Code Generator}. The code generator, represented at the coordinator in Figure~\ref{fig:architecture}, fills pre-defined M4 templates with macros particular to the actual processing requested by the user, generating highly-efficient C++ code similar to direct hard-coding of the processing for the current data. The resulting C++ code is subsequently compiled together with the system code into a dynamic library. This mechanism allows for the execution of arbitrary user code inside the execution engine through direct invocation of the GLA interface methods.

\textit{Code Loader}. The code loader links the dynamic library to the core of the system allowing the execution engine and the GLA manager to directly invoke user-defined methods. While having the code generator at the coordinator is suitable for homogeneous systems, in the case of heterogeneous systems both the code generator and the code loader can reside at the executors.

\textit{Execution Engine}. The EXTASCID execution engine is an enhanced instance of the GLADE-DataPath execution engine. It implements a series of relational operators -- \texttt{SELECT}, \texttt{PROJECT}, \texttt{JOIN}, \texttt{AGGREGATE} -- and a special operator for the execution of arbitrary user code specified using the GLA interface. They are all configured at runtime with the actual code to execute based on the requested processing. The execution engine has two main tasks: manage the thread pool of available processing resources and route data chunks generated by the storage manager to the operators in the query execution plan. Parallelism is obtained by processing multiple data partitions simultaneously and by pipelining data from one operator to another.

\textit{GLA Manager}. The GLA managers at the executors execute the \texttt{Merge} function in the GLA interface~\cite{glade:sigmod}, while the GLA manager at the coordinator executes \texttt{Terminate}. They are dynamically configured with the code to be executed at runtime based on the actual processing requested by the user. Notice that the GLA manager merges only GLAs from different executors, with the local GLAs being merged inside the execution engine.

\textit{Communication Manager}. The communication managers are in charge of transmitting data across process boundaries, between the coordinator and the executors, and between individual executors. Different inter-process communication strategies are used in a centralized environment with the coordinator and the executor residing on the same physical node and for a distributed shared-nothing system. The communication manager at the coordinator is also responsible for maintaining the list of all active executors. This is realized through a heartbeat mechanism in which the executors send \textit{alive} messages at fixed time intervals.

%%%%%%%%%%%%%%%%%%%%%%%%%%%%%%%%%%%%%%%%%%%%%%%%%%%%%%%%%%%%%%%%%%%%%%%%
\eat{
The coordinator as well as the executors consist of multiple components, depicted in Figure~\ref{fig:architecture}. While the components are inherited from the GLADE~\cite{glade:sigmod} architecture, significant enhancements are required to the Storage Manager and GLA Manager modules in order to support extensible array storage and processing in addition to the native relational data model.
}
%%%%%%%%%%%%%%%%%%%%%%%%%%%%%%%%%%%%%%%%%%%%%%%%%%%%%%%%%%%%%%%%%%%%%%%%

%%%%%%%%%%%%%%%%%%%%%
%storage manager
\subsection{Storage Manager}\label{ssec:design:storage}

The storage manager is responsible for organizing data on disk, reading, and delivering the data to the execution engine for processing. There are multiple aspects that distinguish EXTASCID from traditional database storage managers. First, it supports natively relational data as well as multi-dimensional arrays. Second, and most important, the storage manager operates as an independent component that reads data asynchronously and pushes it for processing. It is the storage manager rather than the execution engine in control of the processing through the speed at which data are read from disk. And third, in order to support a highly-parallel execution engine consisting of multiple execution threads, the storage manager itself uses parallelism for simultaneously reading multiple data partitions.

%%%%%%%%%%%%%%%%%%%%%%
\textit{Data partitioning}~\cite{dewitt-paralleldb} represents the main strategy for parallel data processing. In a relational setting, the tuples of a relation are split into multiple segments and assigned to different execution nodes for processing. Since each process works on a considerably smaller dataset, a speedup proportional to the number of processing nodes can be obtained in optimal conditions. In GLADE, data partitioning works as follows. The tuples of a relation are arbitrarily assigned to segments of fixed size---typically a few millions, to increase the size of sequential scans and reduce the number of seeks. This is done at loading by simply following the order in which tuples are ingested. The order of the segments on disk is again arbitrary, typically the order in which they are ingested. The assignment of segments to nodes is round-robin. The goal is to equally divide the data across nodes for load balancing. With this partitioning strategy, queries in GLADE have to always read all the data since there is no relationship between tuples and the segment they are part of. The only reduction in the amount of data read from disk is due to the vertical organization of the segments on disk which allows only for the attributes required by the query to be scanned. When the number of attributes is large, this reduction can be quite significant.

While the GLADE partitioning strategy works for relational data, it is suboptimal for array processing which often requires neighboring cells to be processed together. Specifically, the subsampling \texttt{REBOX} operator can be isolated to the segments overlapping the range selection if data are organized according to their position along the array dimensions. Moreover, the \texttt{APPLY+} operator requires adjacent data to be processed together. If data are not stored organized based on the dimensions, an expensive re-partitioning step is required as pre-processing. The EXTASCID storage manager addresses these problems and optimizes array organization while making minimal modifications to GLADE.

%%%%%%%%%%%%%%%%%%%%%%
\textit{Chunking}~\cite{sarawagi:chunking} or \textit{tiling}~\cite{furtado:tiling} is multi-dimensional range-based data partitioning for parallel array processing. What this means is that data having close values along the set of partitioning attributes are assigned to the same segment. For arrays, dimensions are used as partitioning attributes and the resulting data segments are called chunks or tiles. Possible chunking strategies are presented in~\cite{furtado:tiling,arraystore}. Issues that need to be addressed include the shape of the chunk, the order in which to store the chunks on disk, and how to distribute chunks across processing nodes. Since GLADE already supports data partitioning for relational data, we examine how are these issues addressed for storing array data efficiently in EXTASCID.

The \textit{shape of the chunk} can be fixed across the entire array -- regular chunking -- or there can be multiple shapes, each of them containing the same number of array cells---irregular chunking~\cite{furtado:tiling,arraystore}. Regular chunking is better suited for dense arrays, also known as grids, since each cell in the array contains the same data. The main issue with regular chunking is how to determine the optimal shape. The immediate alternative is to make the size of the chunk along each dimension proportional to the domain size of the corresponding dimension---aligned tiling~\cite{furtado:tiling} uses the same scaling factor across each dimension. Another alternative is to determine the shape based on the query workload as the solution to the optimization formulation that minimizes the overall number of chunks read from disk~\cite{sarawagi:chunking}. Irregular chunking is better suited to sparse arrays. The objective is to create chunks that contain the same number of data points rather than to have chunks with the same shape. This results in similar processing time across chunks and load balancing across processes---an important aspect for parallel processing. EXTASCID supports all these types of chunking. It chooses the appropriate strategy based on the type of data and other available information such as the query workload. Chunking can be executed either as part of the data loading or as an independent process that generates the chunks ready to load in EXTASCID.

Once the chunk shape is determined, two additional problems require attention---\textit{how to order the chunks on disk and how to distribute the chunks across multiple processing nodes}. It is important to notice that no matter what order is chosen, there will be tasks with suboptimal performance. Thus, the idea is to optimize the placement for a given workload or in the average case. Random placement of chunks on disk and across nodes is optimal in the average case. When workload information is available, the order of chunks on disk -- the order in which dimensions are considered -- can be chosen such that chunks that are accessed together are placed contiguously on disk. This results in larger sequential scans and fewer seeks, thus better I/O performance. Larger chunk sizes have a somehow similar effect. The assignment of chunks to nodes involves a more complicated tradeoff. On one side, i.e., subsampling, we aim for maximum parallelism. On the other, i.e., \texttt{APPLY+} operator, the amount of data transferred between nodes has to be minimized. Thus, it is not clear if chunks that are accessed together should be assigned to the same or different nodes. It depends on the actual task to be executed. The problem becomes even more complicated in EXTASCID due to the thread-level parallelism inside each processing node. In this situation, we opt for random chunk placement on disk and random chunk assignment to processing nodes as our default strategy. The user is given the possibility to change this and specify an arbitrary placement though.

%%%%%%%%%%%%%%%%%%%%%
\begin{figure}[htbp]
\centering
\includegraphics[width=0.65\textwidth]{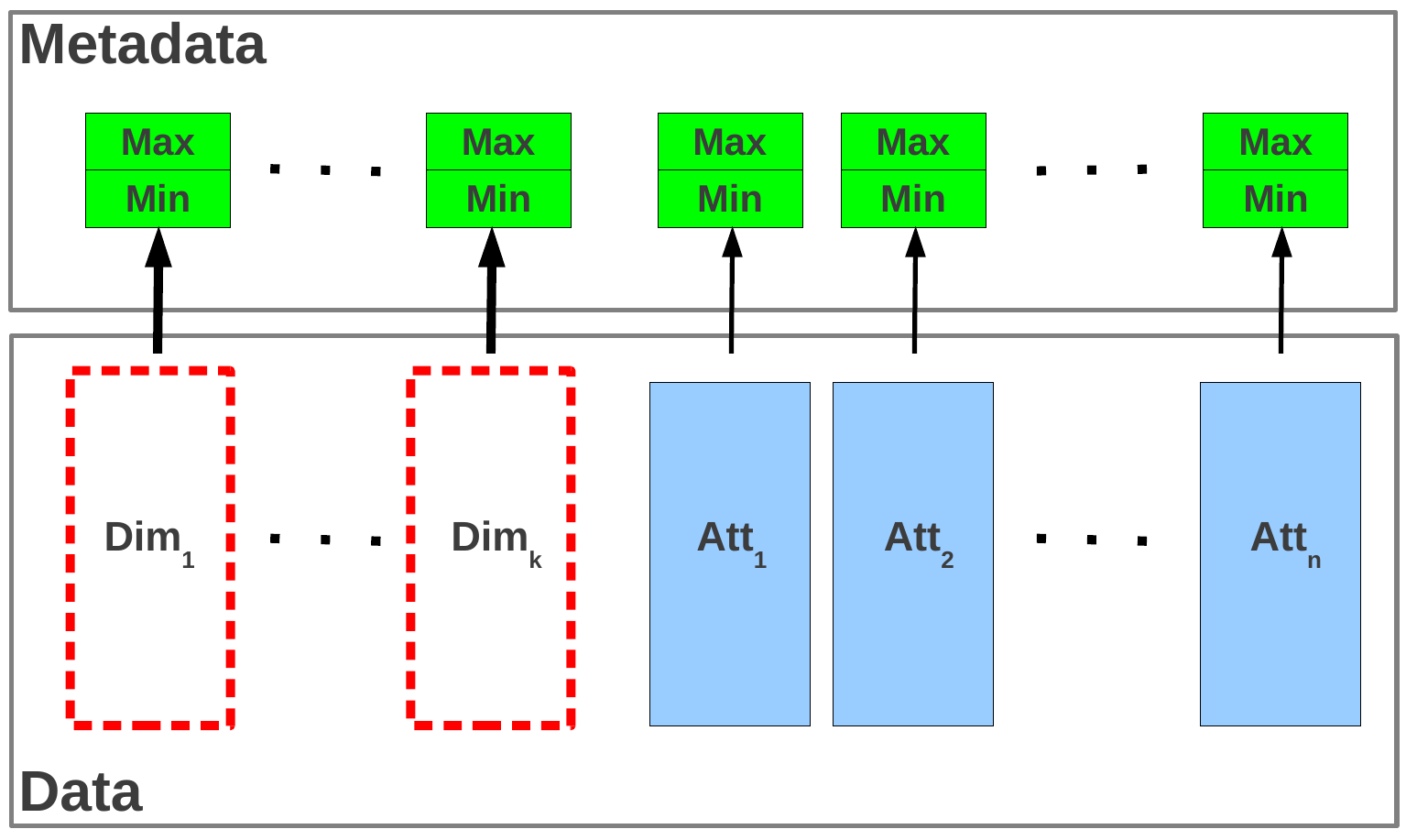}
\caption{Generic chunk structure for the storage of both relational data and arrays in EXTASCID.}
\label{fig:chunk-structure}
\end{figure}
%%%%%%%%%%%%%%%%%%%%%

Figure~\ref{fig:chunk-structure} depicts the \textit{generic structure of an EXTASCID chunk} containing metadata to support range-based data partitioning. It is important to point out that this structure is directly applicable both to relational data as well as arrays. For unordered relations, the dimensions do not exist---there are only attributes. For arrays, dimensions form a key. They have to be represented explicitly for sparse arrays, while in the case of dense arrays the dimensions can be inferred from the position in the chunk when data are stored in a pre-determined order. The metadata contain the minimum and maximum values for each dimension and attribute and are stored in the system catalog. They represent a primitive form of indexing. Different chunking strategies generate different ranges for the (Min, Max) metadata. For example, in the case of regular chunking, the (Min, Max) ranges are equal across all the chunks for all the dimensions and they represent the same fraction from the dimension size. The ranges allow for immediate detection of the chunks that need to be processed in subsampling queries---a large class in array processing. The actual data are vertically partitioned, with each column stored in a separate set of disk blocks. This allows only for the required columns to be read for each query, thus minimizing the I/O bandwidth required for processing. The impact of the (Min, Max) ranges on attributes is not that significant since there is no guarantee that attribute values are clustered. Nonetheless, if there is correlation between the cell position and its value, the (Min, Max) ranges can prune a significant number of chunks even for the value-based \texttt{FILTER} operator.

Given the generic chunk structure, it is important to determine what \textit{optimizations} can be applied for different types of data. We are specifically interested in sparse and dense ordered arrays and unordered relations. While in the case of sparse arrays and relations there is not much beyond using the metadata to determine if a chunk is required for processing in a subsample or selection query, dense arrays provide further optimization opportunities. To be precise, the dimensions can be discarded altogether if the data inside the chunk are stored sorted along a known order of the dimensions. This optimization is known as \textit{dimension} or \textit{index suppression} and can reduce the amount of data read from disk even further. Notice that although index suppression reduces the amount of stored data, we do not consider it as a compression method. Compression is orthogonal to chunk organization. It can be applied at column level. Currently, EXTASCID does not support compression.

As already mentioned before, the EXTASCID storage manager supports a push-based execution model for a merge-oriented parallel execution strategy in which chunks are read from disk asynchronously and injected into the system. This requires the storage manger to determine what chunks have to be generated for each user query. While in GLADE this decision is simple since all the chunks are read for every query, in EXTASCID the chunks required for a subsampling/selection query can be determined based on the (Min, Max) metadata, without actually reading the chunks from disk. This simple form of indexing can result in significant I/O reduction, especially for small range subsampling queries. Following the same strategy of runtime code generation, the storage manager is configured with code to select the chunks based on the query. This pre-processing step is executed during the storage manager setup phase, just before chunks are being generated. The actual process of reading and assembling chunks is highly-parallel, with multiple requests being honored simultaneously. In essence, the storage manager is a complex module consisting of multiple components that operate in parallel and communicate asynchronously.

%%%%%%%%%%%%%%%%%%%%%%
%execution model
\subsection{GLA Execution}\label{ssec:design:execution}

As identified in~\cite{merge-overlap}, there are only two strategies for parallel scientific processing---merge and overlap. In the merge strategy, each data partition is first processed independently by an executor process, followed by a merging phase in which the partial results are combined together. In overlapped execution, enough data are replicated across multiple partitions to isolate any computation to a single data partition, thus eliminating the subsequent merge phase. Merging is a more general strategy, applicable to any computation. Overlapping requires complicated data replication strategies and post-processing and is applicable only to bounded computations---otherwise the entire data have to be replicated at each data partition.

EXTASCID adopts a merge-oriented execution strategy, facilitated by the push-based storage manager and the GLA interface for complex task specification. Merging is supported by two components of the system---a GLA metaoperator that is part of the execution engine and the GLA manager. As all the other operators in the execution engine, the GLA metaoperator takes as input chunks. Unlike other operators though, its functionality is not restricted to a pre-determined template with a reduced number of configuration parameters. Instead, the GLA metaoperator can execute arbitrary user code as long as it is expressed using the GLA interface~\cite{glade:sigmod}. The role of the GLA manager is to merge together GLAs created at different nodes. Merging is executed on arbitrary tree structures, determined independently for each query.

The benefits of the EXTASCID execution strategy are twofold---completeness and extensibility. Since the GLA metaoperator processes chunks, it can handle both relational data and ordered arrays using the same framework. The only difference is the user code which can take advantage of the chunk structure. Extensibility is achieved by executing arbitrary user code expressed using the common GLA interface. In a typical workflow, an operation is first expressed as a GLA and executed by the GLA metaoperator. In time, the operation can be promoted to an independent operator and added to the execution engine.

%%%%%%%%%%%%%%%%%%%%%%
\begin{figure}[htbp]
\centering
\includegraphics[width=0.9\textwidth]{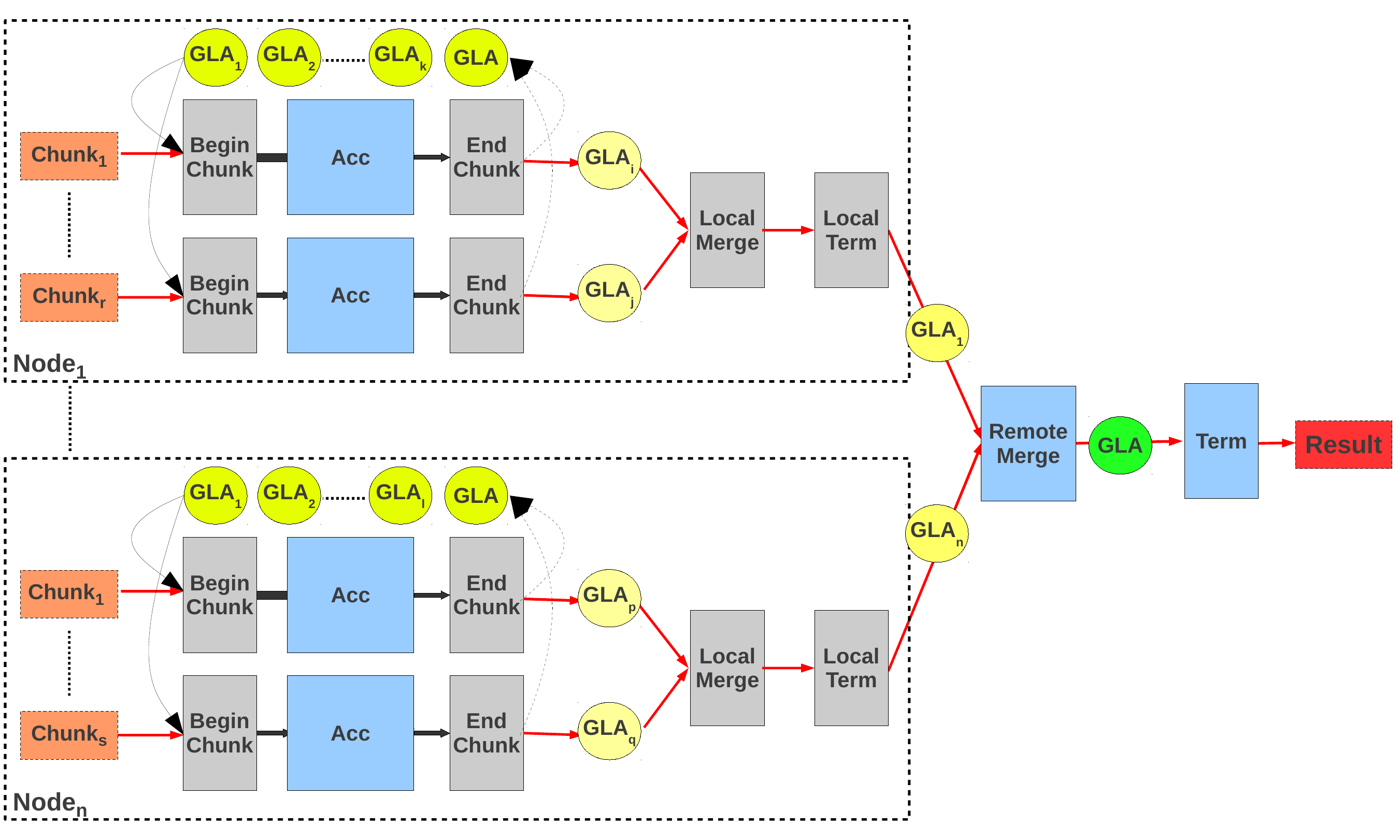}
\caption{EXTASCID merge-oriented execution strategy. The gray rectangles correspond to methods specific to array processing in the extended GLA interface.}
\label{fig:execution}
\end{figure}
%%%%%%%%%%%%%%%%%%%%%%

Figure~\ref{fig:execution} depicts the stages of the merging strategy expressed in terms of the extended GLA interface specific to array processing. The semantic of the standard GLA methods is presented in~\cite{glade:sigmod}. In the following, we focus on the array-specific GLA methods we propose. \texttt{BeginChunk} is invoked before the data inside the chunk are processed, once for every chunk. \texttt{EndChunk} is similar to \texttt{BeginChunk}, invoked after processing the chunk instead. These two methods operate at chunk granularity. They are the places where side-effect operations are executed. For example, data can be sorted according to a dimension that makes the processing more efficient in \texttt{BeginChunk}. In \texttt{EndChunk}, data that are part of the GLA state and do not require further merging can be materialized to disk resulting in significant reduction in memory usage. The difference between \texttt{Init} and \texttt{BeginChunk}, and \texttt{Terminate} and \texttt{EndChunk}, respectively, is that \texttt{BeginChunk} and \texttt{EndChunk} can be invoked multiple times for the same GLA, once for every chunk. This is because GLAs are used across chunks. Merging is invoked in two places. In the GLA metaoperator, \texttt{LocalMerge} puts together local GLAs created on the same processing node, while in the GLA manager \texttt{RemoteMerge} is invoked for GLAs computed at different nodes. This distinction provides optimization opportunities depending on the chunking strategy---when chunks corresponding to the same array are stored on the same node, only \texttt{LocalMerge} is required. \texttt{Terminate} is called after all the GLAs are merged together in order to finalize the computation, while \texttt{LocalTerminate} is invoked after the GLAs at a processing node are merged. \texttt{LocalTerminate} allows for optimizations when the processing is confined to each node and no data transfer is required. It is important to notice that not all the interface methods have to be implemented for every type of processing.

%%%%%%%%%%%%%%%%%%%%%
%array operators implementation
\subsection{Implementation}\label{ssec:extascid:imp}

As mentioned earlier in the paper, EXTASCID is implemented on top of the GLADE~\cite{glade:sigmod} parallel processing system. GLADE in turn uses an extended version of the centralized DataPath~\cite{datapath} relational execution engine for local processing at each node. As a result, EXTASCID inherits the relational algebra operators implemented in DataPath. It also inherits the communication mechanism for parallel aggregation available in GLADE. In this section, we provide more details on how the array-specific ArrayQL algebra operators are implemented in the DataPath execution engine. We consider parallel versions of the operators that process chunked arrays. Chunks are processed independently and in parallel, with minimal data sharing and transfer, in order to maximize the parallelism. Before discussing the array operators, we first look into the parallel implementation of the original relational algebra operators and of the GLA metaoperator.

%%%%%%%%%%%%%%%%%%%%%
%relational operators implementation
\subsubsection{Relational Algebra Operators}\label{ssec:extascid:imp:rel-op}

DataPath~\cite{datapath} implements four operators---\texttt{SELECT}, \texttt{PROJECT}, \texttt{JOIN}, and \texttt{AGGREGATE}. As with any relational database, operators can be combined into execution trees that support the execution of complex queries. The operators process chunks -- a simplified version of the generic chunk structure depicted in Figure~\ref{fig:chunk-structure} -- and generate chunks with a different structure or with different tuples. Inside every operator, the loop iterating over the tuples in the chunk is heavily optimized through compiler specific optimizations such as loop unrolling. This is done for every query in part by generating a hard-coded version of the operator specific to the query, compiling it, and linking it in the execution engine at runtime. The code generation process is driven by a parametrized template specific to each operator that is instantiated with query-specific arguments. This results in code that executes only operator-specific tasks. There is no tuple/attribute packing/unpacking or type conversion and value interpretation.

Specifically, \texttt{SELECT} plugs-in the selection condition in the loop and invalidates the tuples that do not satisfy the condition by resetting a corresponding bit. \texttt{PROJECT} simply drops all the columns in the chunk that are not required further up in the query tree. It is applied for any other operator, including the file-level access methods. \texttt{JOIN} is more complex since there are two stages in the processing---DataPath implements hash join under the assumption that one of the relations fits entirely in memory. In the build phase, the columns corresponding to the small relation are linearized into a hash table based on the join attributes. The hash function call is hard-coded with the specific attributes. In the probing phase, tuples from the large relation are iterated over and matched with corresponding tuples in the hash table to generate result tuples that are vertically partitioned in the chunk structure. Again, all these operations are specific to the query at hand. In \texttt{AGGREGATE}, the function and the arguments -- computed from the chunk attributes -- are the parameters to the code generation template.

%%%%%%%%%%%%%%%%%%%%%
%GLA operator implementation
\subsubsection{GLA Metaoperator}\label{ssec:extascid:imp:GLA}

The GLA metaoperator takes as input chunks. It produces GLAs though. As a result, the GLA metaoperator can be placed only at the root of the query tree in the current implementation. Nonetheless, we can imagine a conversion function that transforms GLAs into chunks, thus allowing the GLA metaoperator to appear anywhere inside a query execution tree. The code generation template for the GLA metaoperator invokes the methods in the extended GLA interface. It is parametrized with the type of the GLA and the expressions the GLA is computed over. It is this additional level of indirection that allows for any user code to be injected into the system, not only valid SQL expressions. To optimize the function call mechanism, the GLA methods are defined inline. While this increases the performance, it also represents a possible source of errors that can bring the entire system down. At the end of the day, the user is allowed to inject any code right in the heart of the system. Our take on this is to aim for maximum performance at the expense of safety and to put the responsibility on the user.

%%%%%%%%%%%%%%%%%%%%%
%array operator implementation
\subsubsection{Array Algebra Operators}\label{ssec:extascid:imp:array}

In order to add new operators to GLADE, we first implement them as special cases of the GLA metaoperator, thus inheriting the parallel merge execution strategy. If the GLA metaoperator is used frequently enough, it can be promoted to become an independent operator with a dedicated keyword and syntactic rules in the query language. We apply this process when we implement the ArrayQL algebra operators in EXTASCID. We present the details for the operators used in the SS-DB benchmark in the following. Before we start though, we emphasize that in the case of sparse arrays -- represented as relations -- no modifications to the relational operators are required.

\texttt{SHIFT} only requires modifications to the chunk metadata since, when stored, the dimension indexes are relative to the chunk origin stored in the metadata. The chunk origin is stored in the global coordinate system.

\texttt{FILTER} is no different from the \texttt{SELECT} operator since the selection is on the values. Unless indexes are defined on the values, the only available solution is to scan all the tuples and check the condition for each of them---this is the DataPath solution. The primitive indexing solution provided in EXTASCID by the (Min, Max) ranges can improve dramatically upon the linear scan when the ranges are tight since a considerably smaller number of chunks have to be read from disk. If there is correlation between the position in the grid and the value, this is the case. \texttt{FILTER} is implemented at two granularity levels. The coarse grained part operates on chunks using the (Min, Max) ranges and is implemented as an access method at the file level. Only the chunks that contain at least a cell satisfying the selection condition are read from disk. The fine grained part is the standard \texttt{SELECT} operator in DataPath. It can be executed independently in parallel for each chunk read from disk.

\texttt{REBOX} is \texttt{FILTER} on dimensions. It follows the \texttt{FILTER} implementation directly, including the parallelism. Given that chunks are clustered on dimensions, it is guaranteed that the minimum number of chunks to be read from disk is always detected in the access method since the (Min, Max) ranges are compact. Moreover, the \texttt{REBOX} operator can take advantage of the chunk organization, i.e., dimension order, and identify the cells in the result box without iterating over all the cells. This represents an optimized version of \texttt{SELECT} for grid data.

\texttt{INNERDJOIN} is a structural join that "glues" together cells at the same index in two different arrays having the same size. If the two arrays are chunked similarly, \texttt{INNERDJOIN} only requires that chunks at the same position are read into memory at the same time. This can easily be enforced in EXTASCID by specifying the same scan order for the two arrays and limiting the number of chunks that are processed inside the execution engine at any time. Chunks at the same index can be processed in parallel independently from all the other chunks. When chunks are non-aligned and have different sizes, \texttt{INNERDJOIN} becomes more complicated since a chunk from one relation can join with many chunks from the other. In the current EXTASCID implementation, \texttt{INNERDJOIN} works only for aligned chunks. Moreover, if one of the arrays fits entirely in memory, the relational \texttt{JOIN} operator can be applied directly.

\texttt{REDUCE} is the exact equivalent of the GLA metaoperator. Thus, it is implemented as specific GLA instances parametrized with the aggregate function. A reduction across a subset of dimensions corresponds to group-by aggregation. It is implemented as a GLA that includes the grouping in all stages of the computation. Parallelism is automatically inherited from the parallel merge strategy specific to the GLA metaoperator.

%%%%%%%%%%%%%%%%%%%%%
%APPLY+ operator implementation
\subsubsection{APPLY+ Operator}\label{ssec:extascid:imp:apply+}

At a high level, \texttt{APPLY+} requires grouping based on a user-defined neighborhood function followed by applying a user-defined aggregate function for all (or a subset of) cells in the array and their corresponding neighbors. The relational representation of \texttt{APPLY+} consists in a structural self-join based on the neighborhood function followed by a group-by aggregation using the user-defined aggregate function. The problem with this two-operator representation is caused by the standard relational \texttt{JOIN} operator which does not consider the ordered array structure. The neighborhood function can be implemented either as a nested-loop join due to the complex join condition or as a series of self-joins--one for each neighbor. Both solutions are inefficient. Consequently, \texttt{APPLY+} is implemented in EXTASCID as a single specialized instance of the GLA metaoperator---the \texttt{APPLY+} GLA.

\texttt{APPLY+} GLA works as follows. In \texttt{BeginChunk}, array cells are sorted such that  they are accessed optimally in \texttt{Accumulate}. Additionally, any other pre-processing required by the neighborhood or aggregate functions, e.g., reinitialize the GLA state, is invoked in \texttt{BeginChunk}. Array cells are grouped according to the neighborhood function in \texttt{Accumulate}, while the aggregate is computed in \texttt{EndChunk}. The simplest implementation of neighborhood grouping is to assign each cell to all the groups it is part of. For SS-DB cooking, this can be further optimized based on the order in which array cells are processed in \texttt{Accumulate}. The completed aggregates can be materialized in \texttt{EndChunk} in order to reduce memory consumption.

Aggregate computation for cells that have neighbors outside of the chunk requires careful consideration. Essentially, the aggregate cannot be computed until all the cells become available in the same GLA. In EXTASCID, this is realized through the parallel merging mechanism -- \texttt{LocalMerge} and \texttt{RemoteMerge} -- provided by the GLA metaoperator. It is important to notice though that the amount of data transferred between GLAs is limited to what is required for the aggregate computation. This is represented exclusively by the GLA state. It is never the case that an entire chunk is passed from one worker node to another if only the border cells are used.

An alternative strategy that avoids merging altogether when the neighborhood function is bounded -- at the expense of increased storage and more complicated chunk management -- is to enforce that aggregate computation is always confined to a chunk, thus allowing for full parallel execution across chunks. This is known as overlapping~\cite{merge-overlap} and requires cell replication across multiple chunks. Multiple overlapping strategies are discussed in~\cite{arraystore}. They differ in the number of cell layers replicated across chunks. In single-layer overlap, a fixed number of border cells from the neighboring chunks are stored together with the chunk---they can also be stored separately and accessed only when required. In multi-layer overlap through materialized overlap views, multiple such layers with increasing thickness are generated and stored. \texttt{APPLY+} GLA can take advantage of overlapping with minimal modifications. The \texttt{Merge} and \texttt{Terminate} methods are not required anymore since the entire computation is finalized in \texttt{EndChunk}. \texttt{BeginChunk} combines the overlapped data with the actual chunk to make them available as a whole in \texttt{Accumulate}.

%%%%%%%%%%%%%%%%%%%%%
%query language
\subsection{Query Language}\label{ssec:extascid:query-lang}

EXTASCID queries are specified at the execution plan level as a query tree that links together the different algebra operators. The plan is written in a query scripting language that requires the specification of the query tree structure -- the operators and the links between them -- and of the functionality of each operator, e.g., the selection ranges for \texttt{REBOX}. While the language might seem not declarative enough, it nonetheless provides a level of abstraction on top of the internal query representation that allows for easy specification of arbitrary user queries---the SS-DB queries have straightforward representations in the language. Essentially, the language is a direct representation for the physical query execution plan. In a full-fledged system with a declarative query language, this form is obtained at the end of a series of transformations and optimizations that take the high-level query to execution. They are all executed transparently to the user. In the current EXTASCID implementation, we bypass this entire process and, since the execution plan is followed exactly, the user is in charge of the query transformation and optimization. In a future implementation, we plan to integrate the upper part of the process starting from ArrayQL~\cite{aql:syntax} or the more powerful SciQL~\cite{SciQL-ideas} language.

%%%%%%%%%%%%%%%%%%%%%
%query execution
\subsection{Query Execution}\label{ssec:extascid:query-exec}

In EXTASCID, query execution starts from the query plan written in the native scripting language. This is passed by the client to the coordinator. The coordinator takes this textual representation and transforms it into an internal format that drives the entire execution. The transformation consists in the instantiation of the generic algebra operators for the given query. The resulting instances behave as hard-coded operators specifically written for the query at hand. There is nothing generic anymore, nothing that requires interpretation. This results in maximum performance since everything is specific to the given query. The exact details of how the entire process works are inherited from the DataPath system and presented elsewhere~\cite{datapath}. The internal query representation is then distributed to all the processing nodes for execution. While this strategy might not be optimal since there are nodes that do not participate in the execution of a given query, e.g., a \texttt{REBOX} operator that accesses chunks from a single processing node, the EXTASCID coordinator cannot decide which nodes to send the query to since it has no global knowledge on data location. The nodes execute the query independently and the partial results are merged on the aggregation tree structure created by the coordinator at query initialization---details on the merging process are presented in~\cite{glade:osr}. If the computation is confined to a chunk or to a processing node -- the case for a large class of array operations -- no merging is required. The extended interface of the GLA metaoperator provides the tools for this type of optimization. The last step of the query execution process is to pass the result to the user. This is done by the coordinator node once merging on the aggregation tree has finalized.

%%%%%%%%%%%%%%%%%%%%%%%%%%%%%%%%%%%%%%%%%
%%%%%%%%%%%%%%%%%%%%%%%%%%%%%%%%%%%%%%%%%
%\input{bench-implement}
\section{Benchmark Implementation}\label{sec:bench-implement}

In this section, we provide the details of how we implement the SS-DB benchmark normal scale, i.e., $400$ grids $7,500 \text{ X } 7,500$ in size, in EXTASCID. We present the structures used to represent the raw images and the derived data -- observations and groups -- the implementation of the array algebra operators used in queries, and the query implementation. We also discuss alternative implementations and the trade-off they incur.

%%%%%%%%%%%%%%%%%%%%%%%%%%%%%%%%%%%%%%
\subsection{Raw Data}\label{ssec:imp:raw}

The raw image array given in Eq.~(\ref{eq:raw-def}) is represented in the local coordinate system. Each image is stored separately and it is chunked regularly into $100$ chunks of equal size $\left(750 \text{ X } 750\right)$. Each chunk is approximately $25\text{MB}$ in size. Grid cells inside the chunk are organized in row-major order without storing the coordinates, i.e., dimension suppression. This results in a $15\%$ reduction in chunk size. Inside the chunk, attributes are stored vertically partitioned and are read only when the query demands it. Each column is approximately $2.25\text{MB}$ in size. The chunks of each image are distributed in round-robin fashion over the available processing nodes. For example, in our experimental setup consisting of $9$ processing nodes with one coordinator and $8$ workers, half of the nodes store $12$ chunks while the other half store $13$ chunks. Overall, each node stores $5,000$ chunks with a total size of $125\text{GB}$. Notice also that each node stores chunks from every image. While other strategies are available for chunk shape, the organization of the cells inside the chunk, and the distribution of chunks across nodes -- see~\cite{array-survey} for a comprehensive survey -- we settled for this approach due to its simplicity and proven efficiency~\cite{arraystore}.

The generic structure shown in Figure~\ref{fig:chunk-structure} stores the chunk borders along each dimension in the chunk metadata. The borders are loaded initially from disk when the chunk is read into memory. During processing, as the chunk passes through operators, the borders are modified accordingly. For example, in order to change to the global coordinate system using the \texttt{SHIFT} operator, chunk borders have to be modified for every chunk in the image based on the origin of the image. This entails only the modification of the borders in the chunk metadata.

%%%%%%%%%%%%%%%%%%%%%%%%%%%%%%%%%%%%%%
\subsection{Derived Data}\label{ssec:imp:derived}

Due to its sparse structure in the global coordinate space, we store derived data in sparse array format. Essentially, dimensions are stored explicitly since their value cannot be inferred anymore from the position in the chunk. In order to support an efficient \texttt{SHIFT} operator, the dimension value is stored relatively to the chunk origin---the minimum value along each dimension stored in the chunk metadata. In the following, we discuss the details specific to the implementation of observations and groups, respectively.

%%%%%%%%%%%%%%%%%%%%%%%%%%%%%%%%%%%%%%
\subsubsection{Observations}\label{ssec:imp:derived:obs}

%%%%%%%%%%%%%%%%%%%%%%%%%%%%%%%%%%%%%%%%%%%%%%%%%%%%%%%%
\eat{

1) How are observations stored in EXTASCID? The relations (or arrays) and their schema.

2) The implementation of the cooking process. What are the functions in the extended GLA interface doing? How are the tables (arrays) corresponding to observations filled? Here we can say that the observations corresponding to each cycle are directed to a single node and they are loaded back into the system as they are generated. One reason to do this is to allow for efficient grouping. Or we can generate a chunk for each image with the local observations at each node. This can be done in EndChunk for non-border observations or in LocalTerminate for the merged observations that are completed. We use different aggregation trees for each image to guarantee that observations on the border are spread uniformly over the nodes. We do not want all the border observations to end-up at the same node. How do we speed-up the merging of border observations? What is the relation between the aggregation tree structure and the distribution of image chunks across nodes? Since we use round-robin, it probably does not matter that much in our case.

3) How is the chunking done at each node? Observations are chunked based on the image they come from. We do not allow observations from different images in the same chunk. Since the number of observations for an image is low (\~20,000), we have a single chunk with observations per image when we redirect all the observations in an image to the same node.

4) How are the chunks corresponding to a group split across nodes? Do we want to have all the observations in a cycle on the same node? I would say NO. Then, after grouping is finished, we take the chunk(s) with observations corresponding to the same cycle and split them across nodes.

}
%%%%%%%%%%%%%%%%%%%%%%%%%%%%%%%%%%%%%%%%%%%%%%%%%%%%%%%%

% array data structures
Observations are represented as two sparse arrays with dimensions explicitly materialized (Table~\ref{tbl:bench-arrays}). \texttt{obs} stores the observation id for every cell that is part of the observation. It has the same dimensionality and coordinate system as \texttt{images}. \texttt{obs\_center} represents the observation center in the global coordinate system and it has as attributes the aggregated properties of the observation, e.g., the average pixel value.

% cooking process
\texttt{obs} is populated by the \texttt{APPLY+} algebra operator during the cooking process. The implementation of the cooking \texttt{APPLY+} operator is standard (Section~\ref{ssec:extascid:imp:apply+}). It has two stages. The first stage identifies the observations internal to an \texttt{images} chunk. It is a fully parallel process without any data transfer. The observations at chunk boundaries have to be merged together in the second stage. This requires transferring the boundary observation data between chunks and even across nodes. Since the number of boundary observations tends to be small, the amount of transferred data is relatively reduced. Moreover, observations are materialized as soon as adjacent chunks are merged in \texttt{RemoteMerge}. \texttt{obs\_center} is populated at the end of the cooking process by combining data from \texttt{images} and \texttt{obs}.

% chunking
There are multiple strategies to chunk \texttt{obs} and \texttt{obs\_center}. For instance, an \texttt{obs} chunk can be created for every corresponding \texttt{images} chunk. The problem with this strategy is that the number of observations in a chunk is too small, e.g., since there are $20,000$ observations on average in an image, less than $200$ observations end-up in the same chunk. Chunks with small size incur reduced I/O throughput due to the frequent disk seeks and short scans. The solution we adopt for increasing the chunk size is to merge observations from multiple chunks together. Merging can be done for the \texttt{images} chunks resident at the same node or, at the extreme, only a single \texttt{obs} chunk is created for every image. After experimenting with these alternatives, we found that having a single \texttt{obs} chunk per image provides optimal performance. This is the solution we implement in EXTASCID. We apply the same chunking strategy, i.e., single chunk per image, for \texttt{obs\_center}. To guarantee uniform distribution of chunks across processing nodes, we build aggregation trees having the root at different nodes while cooking the \texttt{images} array.

%%%%%%%%%%%%%%%%%%%%%%%%%%%%%%%%%%%%%%
\subsubsection{Groups}\label{ssec:imp:derived:groups}

%%%%%%%%%%%%%%%%%%%%%%%%%%%%%%%%%%%%%%%%%%%%%%%%%%%%%%%%
\eat{

1) How are groups stored in EXTASCID? The relations (or arrays) and their schema.

2) The implementation of the grouping process. What are the functions in the extended GLA interface doing? How are the tables (arrays) corresponding to groups filled? Grouping is implemented in parallel only across cycles. Inside the same cycle, we process the chunks corresponding to image observations in order.

}
%%%%%%%%%%%%%%%%%%%%%%%%%%%%%%%%%%%%%%%%%%%%%%%%%%%%%%%%

% array data structures
Groups are represented as two high-dimensional arrays -- \texttt{group\_center} (3-D) and \texttt{group\_center\_img} (4-D) -- with dimensions explicitly materialized. \texttt{group\_center} stores the single group center across all the observations in the group in the global coordinate system. The third dimension corresponds to the cycle the group is part of. \texttt{group\_center\_img} stores a center for every image that contains observations in the group. This center is computed from the observation centers in that particular image. \texttt{img\_id}, taking values between $0$ and $19$, is the fourth dimension in this array.

% grouping process
The only solution to parallelize the grouping process is across different cycles. A merge-based parallel implementation inside the cycle is not possible since all the combinations have to be considered whenever merging observations from multiple images. Essentially, no work is saved through parallelization. As a result, the first step in the grouping process is to bring all the observations corresponding to a cycle, i.e., \texttt{obs\_center}, on the same node. Notice that different cycles end-up on different nodes though. For example, at most 3 cycles from the normal SS-DB instance are processed on the same node in our experimental setup consisting of 8 workers. In terms of the extended GLA interface, all the action happens in \texttt{Terminate}. The global coordinate space (\texttt{img}\_\texttt{id}-\texttt{x}-\texttt{y}) corresponding to a cycle is regularly chunked along the \texttt{x}-\texttt{y} dimensions such that nearby observations are grouped together. The resulting grid index is meant to reduce the number of observations that have to be compared for membership to the same group. Group creation proceeds iteratively from the first image in the cycle with a series of calls to the distance-based \texttt{APPLY+} operator. Once the member observations are determined, the two group arrays -- \texttt{group\_center} and \texttt{group\_center\_img} -- can be filled.

% chunking
We create a single chunk for all the groups contained in a cycle in the \texttt{group\_center} array since the number of groups is at most a constant factor larger than the number of observations in the first image of the cycle. The chunk is obtained directly as a result of grouping. \texttt{group\_center\_img} is sliced additionally along the \texttt{img\_id} dimension, with a chunk generated for every image. Although this requires all the group centers to be inspected for every range query, we have not seen a significant performance degradation when compared to chunking along all the dimensions. The reason is the relatively small number of groups in a cycle.

%%%%%%%%%%%%%%%%%%%%%%%%%%%%%%%%%%%%%%
\subsection{Queries}\label{ssec:imp:queries}

Query implementation follows directly the implementation of the corresponding array algebra operators given in Section~\ref{ssec:ssdb:queries}. The EXTASCID GLA-based implementation for each array operator used in SS-DB is presented in Section~\ref{ssec:extascid:imp:array}. Without going into the details specific to every query in the benchmark, we emphasize two important aspects of our implementation. First, \texttt{REBOX} is pushed for execution into the storage manager. This guarantees that only the chunks required by the query at hand are read from disk, thus minimizing the overall I/O. And second, the chunks that reach the memory are asynchronously processed in parallel both inside the same operator as well as across operators. When combined, these two execution strategies guarantee optimal performance both for range queries as well as for the \texttt{APPLY+} operator, as shown by our experimental results.

%%%%%%%%%%%%%%%%%%%%%%%%%%%%%%%%%%%%%%%%%
%%%%%%%%%%%%%%%%%%%%%%%%%%%%%%%%%%%%%%%%%
%\input{results}
\section{Benchmark Results}\label{sec:results}

In this section, we present the results obtained by executing the SS-DB benchmark in EXTASCID. We use as a relative reference the SciDB implementation made available with the SciDB source code~\cite{scidb}. We optimize the SciDB implementation based on detailed instructions from the SS-DB benchmark maintainers~\cite{ssdb-comm}. We also consider the physical characteristics of the underlying hardware as an absolute reference point for comparison.

%%%%%%%%%%%%%%%%%%%%%%%%%%%%%%%%%%
\paragraph*{System} We report experimental results on a $9$-node shared nothing cluster. Each node has $2$ AMD Opteron $8$-core processors for a total of $16$ cores running at $2\textit{GHz}$, $16\textit{GB}$ of memory, $4$ $1\textit{TB}$ hard-drives, and runs Ubuntu 11.04 $64$-bit. The disks perform sequential reads at $120\textit{MB/s}$ according to \texttt{hdparm}, for $480\textit{MB/s}$ total I/O bandwidth at a node. RAID-0, i.e., striping, is implemented internally in GLADE---no hardware or software RAID controller is part of the system. The nodes are mounted inside the same rack and are inter-connected through a Gigabit Ethernet switch. In EXTASCID, one node is configured as the coordinator while the other $8$ are workers. In SciDB, the coordinator also acts as a processor for a total of $9$ processing nodes. Following the advice of the SS-DB benchmark maintainers~\cite{ssdb-comm}, we use SciDB version 12.7 since the corresponding SS-DB benchmark provided with the source code is stable and performs optimally.

%%%%%%%%%%%%%%%%%%%%%%%%%%%%%%%%%%
\paragraph*{Data} The dataset used in our experiments is the SS-DB normal configuration consisting of $400$ 2-D grids $7,500 \text{ X } 7,500$ in size. The grids are grouped into cycles of $20$, for a total of $20$ cycles. The overall size of the dataset is $1\textit{TB}$ -- each grid is $2.48\textit{GB}$ -- which corresponds to approximately $125\textit{GB}$ allocated to each node. We use the SS-DB configuration with the medium benchmark parameters as defined in~\cite{ssdb}.

%%%%%%%%%%%%%%%%%%%%%%%%%%%%%%%%%%
\paragraph*{Methodology} Benchmark execution consists of three distinct stages. First, raw image data are loaded into the processing system. This is a translation step that maps images from their original representation into the internal system representation. In the second stage, derived data -- observations and groups -- are extracted from the raw data. The overall execution time is reported for each of the three operations---loading, cooking, and grouping. Queries are executed in the third stage as follows. A series of five different configuration parameters are randomly generated for every query. The query is executed ten times for every parameter configuration and the average execution time is reported for the (query, parameter configuration) pair. The same is repeated for every parameter configuration, for a total of five execution times per query. The sum of these execution times is reported as the query execution time. The reason we sum-up the execution time for different configurations is the high variance incurred by different selection ranges, especially in the global coordinate system. We present three results for every benchmark operation---the EXTASCID execution time, the SciDB execution time, and the ratio between the SciDB and the EXTASCID execution time. Notice that all the experiments are executed with cold caches.

%%%%%%%%%%%%%%%%%%%%%%%%%%%%%%%%%%
\subsection{Data Loading}\label{ssec:exp:load}

The results for loading the $400$ benchmark images in EXTASCID and SciDB are presented in Table~\ref{tbl:load-cook-group}. There is a huge difference between the two systems---EXTASCID is faster by a factor of $20$. The main reason for the difference is the parallel loading functionality supported in EXTASCID. The $400$ images are split into groups of $50$, each assigned to a different processing node. Data loading proceeds in parallel across all the nodes. Each image is first chunked and the chunks are round-robin partitioned across all the nodes for loading. Moreover, four different images are processed in parallel at a node, each being read from a separate disk.

In contrast, loading in SciDB is sequential from a single input file. The typical loading process in SciDB consists of two stages. In the first stage, data are ingested into SciDB arrays without chunking which is executed as a reorganization process during the second stage. Interestingly, the SS-DB benchmark implementation in SciDB contains only the first stage, with each image loaded as a single separate chunk of the same array. While supposed to improve query performance, chunking also increases dramatically the loading time---it more than doubles it, to be precise. Notice that the time reported in Table~\ref{tbl:load-cook-group} corresponds to the first loading stage---the only stage implemented in the SS-DB source code~\cite{scidb}.

%%%%%%%%%%%%%%%%%%%%%%%%%%%%%%%%%%
\begin{table}[htbp]
  \begin{center}
    \begin{tabular}{|l||r|r|r|}

    \hline
	\multirow{2}{*}{\textbf{System}} & \multicolumn{3}{|c|}{\textbf{Execution Time [seconds]}} \\ 	

	& Load & Cook & Group \\
	\hline
	\hline
	
	EXTASCID & 1,509 & 82 & 11 \\
	
	\hline
	
	SciDB & 30,361 & 1,086 & 69 \\

	\hline
	\hline
	
	SciDB/EXTASCID & 20.12 & 13.24 & 6.27 \\

	\hline
	
    \end{tabular}
  \end{center}

\caption{Data loading, cooking, and grouping execution time.}\label{tbl:load-cook-group}
\end{table}
%%%%%%%%%%%%%%%%%%%%%%%%%%%%%%%%%%

%%%%%%%%%%%%%%%%%%%%%%%%%%%%%%%%%%
\subsection{Derived Data}\label{ssec:exp:cook-group}

\textbf{Cooking.} The difference between EXTASCID and SciDB -- a factor of $13$ -- is evident even in the case of cooking (Table~\ref{tbl:load-cook-group}). The reasons are different though. EXTASCID performs cooking fully parallel at chunk level using the merge execution strategy. The amount of data transferred between nodes is reduced to the minimum and merging of adjacent chunks is executed as early as possible in order to optimize the performance. SciDB cooking can be executed in parallel only across images due to the original chunking executed during data loading. While it is not clear how a different chunking strategy would perform, we assume this is the optimal SS-DB implementation since it is the only alternative included in the SciDB source code. Optimizing the SS-DB implementation in SciDB is beyond the scope of this work.

\textbf{Grouping.} In the case of grouping, the gap between EXTASCID and SciDB reduces further -- a factor of $6$ -- due to diminishing levels of parallelism available in the EXTASCID implementation. While EXTASCID exploits parallelism only for extracting groups across cycles, the SciDB implementation is entirely sequential. Nonetheless, the two implementations are similar at cycle level. The factor of $6.27$ -- instead of an expected $20$ -- is due to the $8$ nodes available in the experimental system, i.e., at most $8$ cycles can be processed in parallel instead of $20$.

%%%%%%%%%%%%%%%%%%%%%%%%%%%%%%%%%%
\subsection{Queries}\label{ssec:exp:query}

%%%%%%%%%%%%%%%%%%%%%%%%%%%%%%%%%%
\subsubsection{Raw Data}\label{sssec:exp:query:raw}

The results for executing the SS-DB queries on raw data are presented in Table~\ref{tbl:query:raw}. We remark immediately more variability, with EXTASCID still outperforming SciDB at cooking (Q2) and SciDB being slightly faster for Q1 and Q3. We determine the cause for these results in the following. Given the suboptimal chunking of raw data in SciDB -- a single chunk per image -- we execute the queries on the entire image without previously applying any range selection predicate. This is exactly the SS-DB implementation provided with the SciDB source code. SciDB outperforms EXTASCID by a factor of $1.38$ at aggregation over raw data (Q1). While both systems use columnar storage and read only the required data from disk, the difference is made by two features implemented only in SciDB---compression and caching. The amount of data stored on disk for every column is further reduced through lossless compression. This results in shorter I/O delays. In addition, SciDB also implements a chunk-level buffer pool that caches the recently accessed chunks in memory, thus avoiding I/O operations entirely. We observed the effect of caching on Q1, i.e., the first run is four times faster than the subsequent ones. Since Q2 is nothing else than cooking applied to a single image, we expect similar results. This is confirmed in Table~\ref{tbl:query:raw} where we observe that the difference between the two systems is almost the same factor as in the case of cooking the entire dataset. The results for Q3 are almost identical for the two systems. SciDB has a slight advantage over EXTASCID mostly due to compression. Beyond that, the two implementations are identical. It is important to notice that caching does not play a role in this case since Q3 is executed over a full cycle of $20$ images.

%%%%%%%%%%%%%%%%%%%%%%%%%%%%%%%%%%
\begin{table}[htbp]
  \begin{center}
    \begin{tabular}{|l||r|r|r|}

    \hline
	\multirow{2}{*}{\textbf{System}} & \multicolumn{3}{|c|}{\textbf{Execution Time [seconds]}} \\ 	

	& Q1 & Q2 & Q3 \\
	\hline
	\hline
	
	EXTASCID & 22.11 & 8.85 & 129.46 \\
	
	\hline
	
	SciDB & 16 & 110 & 105.24 \\

	\hline
	\hline
	
	SciDB/EXTASCID & 0.72 & 12.43 & 0.96 \\

	\hline
	
    \end{tabular}
  \end{center}

\caption{Execution time for queries on raw data.}\label{tbl:query:raw}
\end{table}
%%%%%%%%%%%%%%%%%%%%%%%%%%%%%%%%%%

%%%%%%%%%%%%%%%%%%%%%%%%%%%%%%%%%%
\subsubsection{Observations}\label{sssec:exp:query:obs}

Table~\ref{tbl:query:obs} contains the results for executing the SS-DB queries over the observation data. EXTASCID is consistently faster than SciDB in this situation, by as much as a factor of $27$ for Q5 and Q6. After careful inspection of the execution mechanisms in the two systems, we found multiple reasons to explain the difference. The root cause is the representation of sparse arrays in the two systems. While EXTASCID supports natively the relational representation of sparse arrays with dimensions stored explicitly, SciDB maps sparse arrays on top of the internal dense array representation which contains entries for all the array cells---valid and invalid. As a result, SciDB has to handle more data and to express queries over sparse arrays on the dense array representation. As the results in Table~\ref{tbl:query:obs} show, this has a negative impact on the execution time. The second cause for the considerably better EXTASCID performance is the suboptimal chunking strategy utilized in SciDB. While this is a direct consequence of the raw data chunking, the effects are considerably worse for observation data since the queries contain range predicates in the global coordinate space. The last reason we found important to explain the gap in execution time is query invocation. In SciDB, each query is executed as $20$ separate queries, one per image in the cycle. In EXTASCID, a single query is sufficient to get the desired result. The iterative invocation of each query and the overhead incurred by the parallel dissemination when accumulated over that many queries become a noticeable fraction in the overall execution time, as the results in Table~\ref{tbl:query:obs} confirm.

%%%%%%%%%%%%%%%%%%%%%%%%%%%%%%%%%%
\begin{table}[htbp]
  \begin{center}
    \begin{tabular}{|l||r|r|r|}

    \hline
	\multirow{2}{*}{\textbf{System}} & \multicolumn{3}{|c|}{\textbf{Execution Time [seconds]}} \\ 	

	& Q4 & Q5 & Q6 \\
	\hline
	\hline
	
	EXTASCID & 0.79 & 0.72 & 0.66 \\
	
	\hline
	
	SciDB & 5 & 20 & 18 \\

	\hline
	\hline
	
	SciDB/EXTASCID & 6.33 & 27.77 & 27.27 \\

	\hline
	
    \end{tabular}
  \end{center}

\caption{Execution time for queries on observation data.}\label{tbl:query:obs}
\end{table}
%%%%%%%%%%%%%%%%%%%%%%%%%%%%%%%%%%

%%%%%%%%%%%%%%%%%%%%%%%%%%%%%%%%%%
\subsubsection{Groups}\label{sssec:exp:query:group}

The results for queries over groups of observations are included in Table~\ref{tbl:query:group}. We remark immediately the high variance in execution time across the two systems considered. While SciDB is faster than EXTASCID by an order of magnitude for Q7 and Q8, EXTASCID outperforms SciDB by more than two orders of magnitude for Q9. Given the similarity of Q8 and Q9, we suspected a problem with the implementation of Q8 in SciDB, confirmed to be true. Essentially, no results are found in Q8 for any of the range predicates, thus no data are retrieved from the raw images. The same happens for some of the ranges in Q9. Whenever data have to be retrieved from the raw images though, the execution time literally explodes. This phenomenon is expected considering that every image is stored as a single chunk. The EXTASCID results on the other hand are always stable and, if we consider the correct execution of Q8, considerably faster. The reason SciDB outperforms EXTASCID on Q7 is because we do not currently store the group centers explicitly. They have to be derived from the member observations every time.

%%%%%%%%%%%%%%%%%%%%%%%%%%%%%%%%%%
\begin{table}[htbp]
  \begin{center}
    \begin{tabular}{|l||r|r|r|}

    \hline
	\multirow{2}{*}{\textbf{System}} & \multicolumn{3}{|c|}{\textbf{Execution Time [seconds]}} \\ 	

	& Q7 & Q8 & Q9 \\
	\hline
	\hline
	
	EXTASCID & 3.36 & 34.25 & 35.87 \\
	
	\hline
	
	SciDB & 0.31 & 2.83 & 3,771.21 \\

	\hline
	\hline
	
	SciDB/EXTASCID & 0.09 & 0.08 & 105.14 \\

	\hline
	
    \end{tabular}
  \end{center}

\caption{Execution time for queries on groups of observations.}\label{tbl:query:group}
\end{table}
%%%%%%%%%%%%%%%%%%%%%%%%%%%%%%%%%%

%%%%%%%%%%%%%%%%%%%%%%%%%%%%%%%%%%
\subsection{Discussion}\label{ssec:exp:discuss}

There are multiple conclusions we can draw from the experimental results presented in this section. The most important point to remark is that although EXTASCID and SciDB share common design and implementation features, their performance is quite different for the SS-DB benchmark. EXTASCID outperforms SciDB by a large margin -- a factor of $10$ on average -- on data loading and derived data computation. This is mostly due to the extensive parallelism available in EXTASCID for this kind of tasks. The relationship between the two systems is more nuanced when considering query performance. SciDB is slightly faster on queries over raw data whenever compression and chunk caching can be applied. For queries over derived data -- amenable to a relational representation -- EXTASCID provides the better execution time since it supports natively both dense as well as sparse arrays. Since SciDB is targeted at dense arrays, mapping between the two representations is required. Overall, the EXTASCID performance is truly remarkable, getting close to the physical bounds imposed by the experimental system for many of the SS-DB benchmark tasks.

%%%%%%%%%%%%%%%%%%%%%%%%%%%%%%%%%%%%%%%%%
%%%%%%%%%%%%%%%%%%%%%%%%%%%%%%%%%%%%%%%%%
%\input{rel-work}
\section{Related Work}\label{sec:rel-work}

There are three lines of research on scientific and array data processing that we consider related to the topics addressed in this paper---benchmarking, array query algebras and languages, and scientific data processing systems. To the best of our knowledge, Sloan Digital Sky Survey (SDSS)~\cite{sdss,sdss:article} is the single other benchmark targeted specifically at scientific processing. Similar to SS-DB, SDSS is also based on processing astronomical images. Unlike SS-DB though, SDSS operates exclusively on relational data obtained as a bi-product of astronomical observations. SS-DB is more general and contains a full spectrum of operations ranging from raw data processing to the creation and querying of derived observation data. These operations manipulate array-oriented data through relatively sophisticated user-defined functions, not always expressible in SQL. While typical implementations of the SDSS benchmark are SQL-based, e.g., MS SQL Server and MonetDB~\cite{monetdb:sky-server}, SciDB and MySQL~\cite{ssdb} are the only two implementations of the SS-DB benchmark we are aware of. With our implementation in EXTASCID, we provide another reference point for a larger adoption of the SS-DB benchmark---the only general benchmark for scientific data processing.

ArrayQL algebra~\cite{aql:algebra} and query language~\cite{aql:syntax} are formalisms introduced in the SciDB context---similar to the SS-DB benchmark. This is the main reason we choose these formalisms to represent the SS-DB operations. They are not commonly accepted though as the representative array algebra and query language. In fact, there is no commonly accepted array algebra and query language. In the following, we discuss other such alternatives proposed in the literature. AQL~\cite{AQL} is a declarative query language for multi-dimensional arrays that treats arrays as functions from index sets to values rather than as collection types. AQL is based on the nested relational calculus with arrays which plays the same role relational calculus and algebra play for the relational data model. The RasDaMan~\cite{baumann:ngits} array algebra conceptualizes arrays as functions from rectangular domains to cell values. Three core constructs -- MARRAY, COND, and SORT -- that can express every array operation when composed together are introduced. In the corresponding RasQL query language~\cite{rasdaman}, arrays are treated as a composite attribute type with a set of corresponding operators. AML~\cite{marathe:arrays} is an algebra consisting of three operators that manipulate dense arrays and take bit patterns as parameters. A significant AML limitation is that it contains only structural operators, i.e., operators that consider the indexes. At query language level, AML is more like an elevated execution plan description than a declarative array query language. RAM~\cite{ram} and SRAM~\cite{sram} are array algebras for dense and sparse arrays, respectively, developed in the context of the MonetDB~\cite{monetdb} columnar database system. Since the execution happens inside a relational database engine, array queries follow a sequence of transformations that take arrays represented in the comprehension syntax to relational operators through an intermediate array algebra stage. Although a series of rewriting rules and optimizations are applied at each of these two steps, relying on the relational algebra operators to map and process array operations introduces inefficiencies due to the impedance mismatch in representation. SciQL~\cite{SciQL-ideas} is the most comprehensive extension to the SQL:2003 standard with support for arrays. It provides seamless integration of set, sequence, and array semantics. The goal is to make minimal modifications to the SQL syntax while allowing for maximum expressiveness in the array operations supported by the language. An interesting characteristic of all the array algebras discussed above is their equivalence. In~\cite{baumann:compare-array-algebra}, it is shown that all the array algebras can be reduced to RasQL---both in array representation as well as operations. This is primarily due to the equivalence between comprehensions and the MARRAY operator for creating arrays. We point the interested reader to~\cite{array-survey} for a comprehensive discussion on these and other array algebras and query languages.

EXTASCID is part of a long series of parallel systems for scientific data processing. Titan~\cite{titan} and T2~\cite{t2} are the first systems designed with extensibility in mind. They adopt an execution strategy closely related to the Map-Reduce~\cite{google:mapreduce} paradigm. More recently, SciHadoop~\cite{scihadoop} implements array processing on top of the popular Hadoop Map-Reduce framework. The main differences between EXTASCID and these systems are the different execution strategy, i.e., UDA vs. Map-Reduce, and the native support for arrays and relations in EXTASCID. RasDaMan~\cite{rasdaman} is a general middleware for array processing with array chunks stored as BLOBs in a back-end database. The processing is specified through a limited number of second-order operators integrated into SQL and executed entirely inside the middleware. RasDaMan is targeted only at array data -- relational data have to be processed either at the application level or inside the middleware -- and it is not parallel. The RAM~\cite{ram} and SRAM~\cite{sram} systems provide support for array processing on top of the MonetDB~\cite{monetdb} columnar database. They do not provide native support for arrays since arrays are represented as relations and array operations are mapped over relational algebra operators. RAM and SRAM are not parallel. SciDB~\cite{scidb:ssdbm-11} is the system EXTASCID resembles the most. Both are parallel systems designed to be extensible. SciDB supports natively only arrays. EXTASCID provides native support both for arrays and relations. The execution strategy in EXTASCID is well-defined through the UDA interface which makes reasoning about parallelism clear. The same is not true in SciDB where a series of UDFs are arbitrarily interconnected. While there are discussions on the implementation of the SS-DB benchmark in multiple of these systems, the SciDB implementation is the only we are aware of at the time when the paper is written. Consequently, this is our reference for the EXTASCID evaluation on the SS-DB benchmark.

%%%%%%%%%%%%%%%%%%%%%%%%%%%%%%%%%%%%%%%%%
%%%%%%%%%%%%%%%%%%%%%%%%%%%%%%%%%%%%%%%%%
%\input{conclusions}
\section{Conclusions}\label{sec:conclusions}

In this paper, we present a formal representation of the SS-DB benchmark in terms of array algebra operators and array query language constructs. These are meant to simplify the implementation in other systems and foster the acceptance of SS-DB as the standard benchmark to evaluate scientific data processing applications. Given that no alternatives exist, SS-DB fills an important void in the evaluation of a large class of Big Data applications. To verify the soundness of our formalization, we give a reference implementation and present benchmark results in EXTASCID, a novel system for scientific data processing we have developed. EXTASCID is complete in providing native support both for array and relational data and extensible in executing any user code inside the system by the means of a configurable metaoperator. These features result in an order of magnitude improvement over SciDB at data loading, extracting derived data, and operations over derived data. The results prove that the integrated EXTASCID architecture supporting natively both arrays and relations is more suited for complex scientific processing over raw and derived data requiring a high degree of extensibility.

%%%%%%%%%%%%%%%%%%%%%%%%%%%%%%%%%%%%%%%%%
%%%%%%%%%%%%%%%%%%%%%%%%%%%%%%%%%%%%%%%%%
\bibliographystyle{abbrv}
%\bibliography{biblio}

\end{document}